\documentclass[12pt]{article}
\pdfoutput=1
\usepackage{epsf}
\usepackage{amsmath}
\usepackage{amsfonts}
\usepackage{amssymb}
\usepackage{graphicx}
\usepackage{color}
\usepackage{psfrag}
\usepackage{cite}
\usepackage{subcaption}
\usepackage{hyperref}
\usepackage{tikz}

\usepackage{ifpdf}

\newcommand{\bmat}{\left(\begin{array}}
\newcommand{\emat}{\end{array}\right)}

\def\yzero{\smash{\hbox{$y\kern-4pt\raise1pt\hbox{${}^\circ$}$}}}

\def\beq{\begin{equation}}
\def\eeq{\end{equation}}
\def\beqa{\begin{eqnarray}}
\def\eeqa{\end{eqnarray}}

\def\-{\hphantom{-}}

\def\s2{\frac{1}{\sqrt2}}

\def\beq{\begin{equation}}
\def\eeq{\end{equation}}
\def\beqa{\begin{eqnarray}}
\def\eeqa{\end{eqnarray}}
\def\tr{{\rm tr \,}}

\def\IF{\relax{\rm I\kern-.18em F}}
\def\II{\relax{\rm I\kern-.18em I}}

\def\Dsl{\,\raise.15ex\hbox{/}\mkern-13.5mu D} 

\def\IZ{{\bf Z}}

\def\NN{{\cal N}}

\newcommand{\eq}[1]{(\ref{#1})}
\newcommand{\ket}[1]{\vert #1 \rangle}

\catcode`\@=11   
\newdimen\@rotdimen
\newbox\@rotbox  

\def\@vspec#1{\special{ps:#1}}
\def\@rotstart#1{\@vspec{gsave currentpoint currentpoint translate
   #1 neg exch neg exch translate}}
\def\@rotfinish{\@vspec{currentpoint grestore moveto}}
\def\@rotr#1{\@rotdimen=\ht#1\advance\@rotdimen by\dp#1%
   \hbox to\@rotdimen{\hskip\ht#1\vbox to\wd#1{\@rotstart{90 rotate}%
   \box#1\vss}\hss}\@rotfinish}
\def\@rotl#1{\@rotdimen=\ht#1\advance\@rotdimen by\dp#1%
   \hbox to\@rotdimen{\vbox to\wd#1{\vskip\wd#1\@rotstart{270 rotate}%
   \box#1\vss}\hss}\@rotfinish}%
\def\@rotu#1{\@rotdimen=\ht#1\advance\@rotdimen by\dp#1%
   \hbox to\wd#1{\hskip\wd#1\vbox to\@rotdimen{\vskip\@rotdimen
   \@rotstart{-1 dup scale}\box#1\vss}\hss}\@rotfinish}%
\def\@rotf#1{\hbox to\wd#1{\hskip\wd#1\@rotstart{-1 1 scale}%
   \box#1\hss}\@rotfinish}%
\def\rotate{\@ifnextchar[{\@rotate}{\@rotate[l]}}
\def\@rotate[#1]#2{\setbox\@rotbox=\hbox{#2}\@nameuse{@rot#1}\@rotbox}

\catcode`\@=12

\topmargin
-1.5cm
\textwidth
15.5cm
\textheight
23.5cm
\oddsidemargin
0.7cm
\evensidemargin
1.2cm

\begin{document}

\makeatletter
\@addtoreset{equation}{section}
\makeatother
\renewcommand{\theequation}{\thesection.\arabic{equation}}
\hypersetup{pageanchor=false}
\pagestyle{empty}
\rightline{ IFT-UAM/CSIC-17-012}
 \rightline{MPP-2017-26}
\vspace{1.2cm}
\begin{center}
\LARGE{\bf A Chern-Simons Pandemic \\[12mm]}
\large{Miguel Montero$^{1}$,  Angel M. Uranga$^2$, Irene Valenzuela$^{1,3}$\\[4mm]}
\footnotesize{${}^1$Institute for Theoretical Physics and
Center for Extreme Matter and Emergent Phenomena,\\
Utrecht University, Princetonplein 5, 3584 CC Utrecht, The Netherlands\\
${}^2$ Instituto de F\'{\i}sica Te\'orica IFT-UAM/CSIC,\\[-0.3em] 
C/ Nicol\'as Cabrera 13-15, 
Campus de Cantoblanco, 28049 Madrid, Spain \\ 
${}^3$ Max-Planck-Institut fur Physik,
Fohringer Ring 6, 80805 Munich, Germany\\
}

\vspace*{1.5cm}

\small{\bf Abstract} \\[5mm]
\end{center}
\begin{center}
\begin{minipage}[h]{16.0cm}
In this paper we study the consistency of generalized global symmetries in theories of quantum gravity, in particular string theory. Such global symmetries arise in theories with $(p+1)$-form gauge fields, and for spacetime dimension $d\leq p+3$ there are obstructions to their breaking even by quantum effects of charged objects. In 4d theories with  a 2-form gauge field (or with an axion scalar), these fields endow Schwarzschild black holes with quantum hair, a global charge leading to usual trouble with remnants. We describe precise mechanisms, and examples from string compactifications and holographic pairs, in which these problems are evaded by either gauging or breaking the global symmetry, via (suitable versions of) Stuckelberg or Kaloper-Sorbo couplings. We argue that even in the absence of such couplings, the generic solution in string theory is the breaking of the global symmetries by cubic Chern-Simons terms involving different antisymmetric tensor fields. We conjecture that any theory with (standard or higher-degree antisymmetric tensor) gauge fields is in the Swampland unless its effective action includes such Chern-Simons terms. This conjecture implies that many familiar theories, like QED (even including the charged particles required by the Weak Gravity Conjecture) or $\NN=8$ supergravity in four dimensions, are inconsistent in quantum gravity unless they are completed by these Chern-Simons terms.

\end{minipage}
\end{center}
\newpage
\setcounter{page}{1}
\pagestyle{plain}
\renewcommand{\thefootnote}{\arabic{footnote}}
\setcounter{footnote}{0}

\setcounter{tocdepth}{2}
\hypersetup{pageanchor=true}
\tableofcontents

\vspace*{1cm}
\section{Introduction}
Global symmetries are a powerful tool in Quantum Field Theory, yet they have convincingly been argued to be incompatible with theories including Quantum Gravity\cite{Abbott:1989jw,Coleman:1989zu} (see \cite{Kallosh:1995hi,Banks:2010zn} for recent discussions) \footnote{For alternative viewpoints, see \cite{Dvali:2012rt,Dvali:2016mur}.}, most notably String Theory \cite{Banks:1988yz}. In fact, the absence of exact global symmetries is the prototype of a set of ideas distilling out the Landscape from the Swampland \cite{Vafa:2005ui}, i.e. effective field theories which cannot be embedded in a theory of Quantum Gravity (see e.g. \cite{Ooguri:2006in,Adams:2006sv} for additional ideas, in particular the Weak Gravity Conjecture \cite{ArkaniHamed:2006dz} and related recent activity \cite{delaFuente:2014aca,Rudelius:2014wla,Rudelius:2015xta,Montero:2015ofa,Brown:2015iha,Bachlechner:2015qja,Hebecker:2015rya,Brown:2015lia,Junghans:2015hba,Palti:2015xra,Heidenreich:2015nta,Kooner:2015rza,Heidenreich:2015wga,Ibanez:2015fcv,Montero:2016tif,Heidenreich:2016aqi,Hebecker:2016dsw,Saraswat:2016eaz,Herraez:2016dxn,Ooguri:2016pdq,Cottrell:2016bty,Hebecker:2017wsu}).

Most discussions about global symmetries in Quantum Gravity focus on symmetries under which charged objects are particles. The main argument supporting the violation of global symmetries involves evaporation of black holes by emission of charged particles, with some plausible assumption about the number of possible remnants not getting to large in a consistent gravity theory  \cite{Susskind:1995da,Banks:2010zn} (in string theory, there are stronger arguments at the perturbative level \cite{Banks:1988yz}, or in holographic setups \cite{Beem:2014zpa}). 

In this work, we show that arguments along those lines can also be applied to generalized global symmetries, in the sense of \cite{Gaiotto:2014kfa}. A prototypical example arises in theories with $(p+1)$-form gauge fields $C_{p+1}$, which in the absence of  charges have fields strengths $F_{p+2}$ obeying $dF_{p+2}=0$, $d*F_{p+2}=0$; they lead to generalized global symmetries with conserved currents $j=F_{p+2}$, $j'=*F_{p+2}$, with charged operators given by (exponentials of) generalized Wilson lines of the electric and magnetic gauge potentials on non-trivial cycles. In high enough dimension, the existence of charged objects breaks these symmetries explicitly even in the vacuum e.g. via loop-effects, but in lower dimensions IR effects confine charged objects and prevent this kind of breaking. Concretely, this applies to the shift symmetry of a 2d periodic axion-like scalar $\phi$, for which the global symmetry is associated to the current $j=d\phi$. In four spacetime dimensions, which is our main focus, this occurs for the dynamics of a 4d 2-form gauge field; we actually show the presence of this fields can endow Schwarzschild black holes with quantum hair, which manifests as a charge under the corresponding 2-form global symmetry, leading to troubles with remnants. This puts the incompatibility of generalized global symmetries with Quantum Gravity on a similar footing to usual global symmetries.

The conclusions generalize to a $(p+1)$-form gauge potential $C_{p+1}$ in spacetime dimension $d=p+3$.
This can be used to study Quantum Gravity constraints on theories with (arbitrary-rank antisymmetric tensor) gauge fields. Such constraints can be extended even to higher dimensions, if we assume the theory should make sense upon compactification. In particular, in principle any theory with a $(p+1)$-form gauge field in $D$ dimensions is in danger of falling into in the Swampland, since it can be compactified down to $d=p+3$ and lead to problematic global symmetries.

There are several ways in which a theory may escape this fate within field theory, which roughly amount  either to promote the global symmetry to a gauge symmetry, or to break it. To introduce them, note that the symmetry acts by shifting $C_{p+1}\to C_{p+1}+\Lambda_{p+1}$ with $\Lambda_{p+1}$ a closed form. Illustrative examples of the two mechanisms are:

\medskip

$\bullet$ Gauging it by introducing a $(p+2)$-form gauge potential $C_{p+2}$ transforming as $C_{p+2}\to C_{p+2}+d\Lambda_{p+1}$, for general $\Lambda_{p+1}$; the global symmetry thus simply becomes the global part of a gauge symmetry, obtained by restricting to closed $\Lambda_{p+1}$. An example is the removal of the global shift symmetry for a 4d axion $\phi$  by introducing a standard $U(1)$ gauge field $A$ with lagrangian $|d\phi-A|^2$. Another example is the gauging of the symmetry associated to a 2-form field $b_2$ by coupling it to a 3-form $c_3$ with lagrangian $|db_2-c_3|^3$, as in the dual description \cite{Marchesano:2014mla} of axion monodromy \cite{Silverstein:2008sg,McAllister:2008hb}.

\medskip

$\bullet$ Breaking it by coupling it to a $(d-p-2)$-form gauge potential $C_{d-p-2}$ via a $d$-dimensional coupling $C_{p+q}\, F_{d-p-1}$, with $F_{d-p-1}=dC_{d-p-2}$. A prototypical example is the breaking of the shift symmetry of a 4d axion by coupling it to a 4-form field strength $\phi F_4$; as in the Kaloper-Sorbo description \cite{Kaloper:2008fb} of axion monodromy \footnote{See also related realizations involving gauge instantons \cite{Dvali:2005an,Dvali:2005ws}, or D-brane instantons in string theory \cite{Garcia-Valdecasas:2016voz}.}. Another example is given by $BF$ Stuckelberg couplings in 4d, in which coupling to a standard $U(1)$ gauge field breaks the global symmetry shifting periods of the 2-form $B_2$.

\medskip

It is interesting, and clear from the above examples,  that considering the symmetries associated to a gauge field and its magnetic dual, the gauging of one is equivalent to breaking of its dual, and vice-versa. It is also interesting that both mechanisms cannot be present for the same $p$-form. For instance, a 4d 2-form $b_2$ cannot have simultaneously a $b_2F_2$ Stuckelberg coupling if its dual axion $\phi$ has a $\phi F_4$ Kaloper-Sorbo coupling, since together they combine into a Green-Schwarz-like mechanism rendering the theory not gauge invariant. In string theory realizations, this follows from microscopic consistency conditions \cite{BerasaluceGonzalez:2012zn} \footnote{\label{foot-gs} There is a mixed 't Hooft anomaly (from a Green-Schwarz diagram) which prevents us from gauging the two dual potentials at the same time \cite{Gaiotto:2014kfa}; for $F_4=\tr F^2$ for some non-abelian group, both can be present if there are additional charged matter fermions contributing to the anomaly; the fermion phase contributes another scalar, which mixes with the axion in two orthogonal combinations, one with Stuckelberg coupling and one with Kaloper Sorbo coupling. }.

\medskip

The above topological couplings are often present in string vacua. For instance, in 4d string compactifications,  $BF$ couplings play a fundamental role in making certain $U(1)$ gauge fields massive (in particular, anomalous ones), see e.g. \cite{Ibanez:1998qp,BerasaluceGonzalez:2011wy}; also $\phi F_4$ couplings underlie stabilization of axion components in moduli \cite{Marchesano:2014mla,McAllister:2014mpa}. On the other hand, there are many examples in which such couplings (or their analogues in other dimensions) are {\em not} present. In fact this is the case even for many supersymmetric AdS vacua with well-defined holographic duals, including the celebrated type IIB on AdS$_5\times S^5$.  These string compactifications and  constructions would seemingly suffer from the generalized global symmetry problem, yet they correspond to consistent string theory configurations. 

The solution to this conundrum is that string theory provides a slightly more subtle (yet related) mechanism to break the symmetries \footnote{As we will discuss in Section \ref{sec:holex}, there is an explicit solution in the AdS$_5\times S^5$ example involving stringy effects. However, even in this example, the mechanism described in the main text is present.}, as follows. Even in the absence of the above quadratic couplings, string models  contain cubic Chern-Simons couplings, e.g. those arising upon reduction of the 10d supergravity couplings $H_3\, F_p\, F_{7-p}$ among NSNS and RR forms in type II, I theories, and generalizations (or heterotic dual versions) thereof. In the presence of internal fluxes in compactification to e.g. four dimensions, the  4d theory displays quadratic $BF$ or $\phi F_4$ couplings, bringing us to the solutions described above. On the other hand, in the absence of internal flux, there is no actual quadratic coupling, and the 4d effective theory seems to describe a massless gauge field with a global symmetry problem. However, the presence of the cubic Chern-Simons term also produces the breaking of the symmetry, even in the fluxless vacuum: The theory contains domain walls, that  can be used to nucleate bubbles inside which the flux is turned on; effects of virtual bubbles of this kind nicely illustrate the breaking of the symmetry. The argument easily generalizes to arbitrary numbers of dimensions. Therefore, the generalized global symmetry problem of string compactifications with (possibly antisymmetric-tensor) gauge fields is solved by the presence of cubic Chern-Simons terms (we also provide additional support from considerations of gauge-gravity holography). Conversely, this provides a partial explanation for the presence of these terms in the string effective action.

From this latter viewpoint, we can promote our observations to a conjecture, in the landscape/swampland spirit. Any field theory with (possibly antisymmetric tensor) gauge fields is not compatible with Quantum Gravity if it does not include Chern-Simons terms breaking the corresponding generalized global symmetries\footnote{Topological Chern-Simons terms have played an important role (albeit seemingly unrelated to ours) in outcasting to the Swampland certain 10d $\NN=1$ supergravities with groups $U(1)^{496}$ and $E_8\times U(1)^{248}$ \cite{Adams:2010zy}.}. 
We provide examples of seemingly reasonable theories which do not survive it, for instance $U(1)$ gauge theory even after the inclusion of the charged particles required by the Weak Gravity Conjecture, pure gravity in $d\geq 4$ dimensions, or four-dimensional $\mathcal{N}=8$ SUGRA.

We have learned about the work \cite{Hebecker-Soler}, which focuses on the black holes with $B$-field quantum hair which were our original motivation, and studies the dynamics of their evaporation to obtain new cutoffs on the effective field theory. The present paper focuses on the generic mechanism by which the problems with this and similar situations seem to be resolved in string theory.

The paper is organized as follows: Section \ref{sec:rev:GGS} provides a brief introduction to the aspects of generalized global symmetries which will be useful later on. Section \ref{sec:BGHHS} discusses a Schwarzschild black hole with $B$-field quantum hair, showing that an ungauged/unbroken symmetry leads to a remnant problem. Section \ref{sec:break} explores the different ways to solve this problem, either by gauging or breaking the symmetry via different mechanisms. Section \ref{sec:conj} presents our conjecture and discusses implications and rationale for it. Section \ref{sec:examples} presents examples in support of the conjecture. Section \ref{sec:3forms} extends the conjecture to $(d-1)$-form fields in $d$ dimensions. Finally, we present our conclusions in Section \ref{sec:conclus}.

We have tried to structure the paper in such a way that the different Sections are as self-contained as possible. Readers interested mostly in black holes with $B$-field hair can focus on Sections \ref{sec:BGHHS} and \ref{sec:break}. Those who care mostly about the conjecture, implications, and ``experimental'' support, can read from Section \ref{sec:break} on, particularly Sections \ref{sec:conj} and \ref{sec:examples}. Section \ref{sec:rev:GGS} is a review which includes no new material.

\section{A brief review of generalized global symmetries}\label{sec:rev:GGS}
We will start by reviewing a few aspects of generalized global symmetries which will help us organize and homogenize the discussion in the rest of the paper. The reader already familiar with \cite{Gaiotto:2014kfa} may safely skip this Section. 

Ordinary symmetries are described by a parameter $\lambda$ (we will only be concerned with abelian rank 1 symmetries). This parameter generates a one-parameter family of transformations on fields, such that expectation values are invariant. It is very natural to think of $\lambda$ to vary from point to point, even if the symmetry is global (for instance, this is how Ward identities are derived, see e.g. \cite{Polchinskiv1}). Since local operators are inserted at a point, namely on a 0-cycle, the infinitesimal action of the symmetry can be recast in terms of the natural pairing between cycles and forms,
\begin{align}\frac{\delta_\lambda \phi(x)}{\phi(x)}=\lambda(x)=\int_x \lambda.\end{align}
Using Hodge duality, we might as well describe the global symmetry with the $d$-form Hodge dual parameter $*\lambda$. 

Generalized global symmetries are constructed by taking the symmetry parameter to be an arbitrary closed $p$-form, $\lambda_p$. These symmetries no longer act in a natural way on local operators, but rather on operators $\mathcal{O}_{A_p}$ defined on $p$-cycles $A_p$, so that
\begin{align}\delta \mathcal{O}_{p}=\left(\int_{A_p}\lambda_p\right) \mathcal{O}_p.\label{genact}\end{align}
Thus, correlators are invariant under infinitesimal shifts of the $\mathcal{O}_{p}$. 

Noether's theorem relates ordinary global symmetries with conserved currents, 1-forms satisfying $d*j=0$. There is  a similar relationship between generalized $p$-form global symmetries and $(p+1)$ conserved currents, $d*J_{p+1}=0$.  

Gauging the symmetry amounts to dropping the requirement that $\lambda_p$ must be closed, while demanding that \eq{genact} is still a symmetry. In the $0$-form example this amounts to the familiar condition that symmetry parameter is now allowed to be nonconstant. In this case, invariance demands the introduction of an additional $1$-form potential, and charged local operators must be dressed by a gauge field Wilson line along a path ending at the insertion point. The story is similar in the generalized case; one introduces a local $p+1$-form potential $C_{p+1}$, which transforms by $d\lambda_p$, and also demands that correlators including $\mathcal{O}_{A_p}$ vanish unless they also contain a factor 
\begin{align}\int_{\mathcal{A}_{p+1}}C_{p+1}\label{gaugingggs}\end{align}
where $\partial \mathcal{A}_{p+1}=A_p$. The original generalized global symmetry still survives, as the subgroup with $\lambda_p$ closed. Because of the coupling \eq{gaugingggs}, the operator $\mathcal{O}_{A_p}$ must actually be a singlet under transformations with closed $\lambda_p$, unless $\mathcal{A}_{p+1}$ extends all the way to infinity. In compact spaces this cannot happen, and thus every correlator is forced to be invariant under the original generalized global symmetry; this is Gauss' law.

We can alternatively  break the symmetry explicitly. In the $0$-form case, this means that the divergence of the current is now nonvanishing:
\begin{align} d*J\neq0\end{align}
A similar statement holds for higher symmetries: The current $J_{p+1}$ has vanishing divergence $d*J_{p+1}$. As an illustration, consider QED in four dimensions. There is a putative 1-form global symmetry, which acts by shifting the periods of the electromagnetic potential $A$. The charged operators are Wilson loops, since
\begin{align} \exp\left(i\int_{C} A\right)\rightarrow  \exp\left(i\int_{C} \lambda_1\right) \exp\left(i\int_{C} A\right),\end{align}
that is, they are multiplied by a phase. The associated current is just $F$, and the conservation equation is $d*F=0$. This symmetry is broken explicitly by the introduction of electrically charged particles, since we now have $d*F=j_e$, with $j_e$ the electric current. 

Generalized gauge symmetries lead to lower-degree generalized symmetries under compactification. For instance, the 1-form symmetry of QED we just discussed leads to one $0$-form and one $1$-form symmetry when compactifying on $S^1$ (plus two other symmetries coming from the magnetic potentials). The 3d $1$-form symmetry is the same 4d $1$-form symmetry, restricted to the three-dimensional gauge field, and the $0$-form symmetry consists of shifts of the three-dimensional axion $\phi=\int A_4$. From this perspective, it is clear that charged objects in the higher-dimensional theory break the symmetry: Electric particles running on the $S^1$ become instantons for the three-dimensional axion $\phi$, breaking its shift symmetry.

Perhaps the most familiar examples of generalized global symmetries  in string theory is given by $p$-form gauge potentials from RR and NSNS fields (or heterotic counterparts). Focusing on e.g. the RR potentials $C_{p+1}$, they enjoy a gauge invariance $C_{p+1}\rightarrow C_p+d\lambda_{p}$. The charged operators are associated to D-branes. A closed $\lambda_{p}$  which does not vanish at infinity generates a symmetry of the theory, in which the D-branes get a phase given by their couplings 
\begin{align}\int_{\text{D-brane}}C_{p+1}.\label{dbraneRRcop}\end{align} This symmetry is gauged, and so is exact. On top of this, the supergravity action is also invariant under the ungauged, $(p+1)$-form global symmetries $C_{p+1}\rightarrow C_{p+1}+ \Lambda_{p+1}$, where $\Lambda_p$ is any closed $p$-form, not necessarily exact. These symmetries are explicitly broken by the D-branes themselves, since their couplings \eq{dbraneRRcop} are manifestly not invariant under the symmetry. Another way of saying this is that the associated currents $*F_{p+2}$ have nonvanishing divergence, as explained above. The case when $p=d'-3$, where $d'$ is the number of noncompact dimensions, requires further consideration, and will be the main subject of this paper, as explained in Section \ref{sec:unbroken}.

Furthermore, if $d>p+3$, generalized global symmetries can be spontaneously broken in the vacuum. For instance, in the above 4d QED  example, the global 1-form symmetry is spontaneously broken, and the photon is precisely the Goldstone mode. This in turn forces the low-energy lagrangian to be the standard Maxwell term. However, we will also be concerned  with theories with $d\leq p+3$, in which the Coleman-Mermin-Wagner prevents spontaneous breaking of the symmetry.

\section{Black holes with generalized global symmetry \texorpdfstring{\\}{ } charge}
\label{sec:BGHHS}

In this section we consider generalized 2-form global symmetries in 4d, and explore their role in the physics of Schwarzschild black holes. The purpose is two-fold: on one hand, this analysis explains the physical black hole arguments underlying the compatibility problems of generalized global symmetries with quantum gravity, placing then on similar footing with the problems of usual global symmetries. On the other hand, the Euclidean version of the system can be regarded as a prototype of a 2d compactification, providing a good template for other generalized global symmetries. The main discussion for the latter are anyway recapped and extended in Section \ref{sec:break}, to which the interested reader may jump safely.

\subsection{Global hair for black holes}
\label{sec:hair}

We will now address the problem that originally motivated this work. It is a widely accepted folklore theorem that theories of quantum gravity do not have exact global symmetries\footnote{See however \cite{Dvali:2012rt,Dvali:2016mur} for alternative viewpoints.}. There are several loose arguments for this. For instance \cite{Susskind:1995da,Banks:2010zn} , if we had such a symmetry, we would be able to build way too many (actually infinitely many) charged black hole states, which would lead to a pathology at arbitrarily low energies. For instance, the energy density due to the Unruh effect for a uniformly accelerated observer in flat space would diverge. 
 
Although the argument is not to be taken too seriously, it does tell us that we should be wary of any theory in which one can endow black holes with ``global hair'', which does not backreact on the metric. In the following we consider one such realization. We will focus on a simple four-dimensional theory which includes gravity, plus a dynamical 2-form field $B_2$. Such systems are ubiquitous in string theory: $B_2$ can arise directly from dimensional reduction from the higher $p$-form potentials, or as the four-dimensional dual potential to an axion $\phi$. The latter arises in particular in any axion inflation model (although the standard inflationary description in terms of the axion does not make it manifest). The relevant part of the action is roughly of the form
\begin{align}\int \frac 12\,\vert H_3\vert^2,\quad H_3=dB_2\end{align}
where by ``roughly'' we mean that the kinetic term coefficient can depend on other fields, such as the dilaton or other moduli of the compactification.
 
This system enjoys a symmetry given by 
\begin{align}B_2\rightarrow B_2+\Lambda_2,\label{2formgs}\end{align}
 where $\Lambda_2$ is any closed 2-form. This is an example of 2-form global symmetry, as discussed in Section \ref{sec:rev:GGS}:  If the 4d spacetime $X_4$ contains noncontractible 2-cycles $\Sigma$, we can compute the periods
 \begin{align}\int_\Sigma B_2\end{align}
 which are higher-dimensional versions of Wilson lines. In analogy to their their $0$-form counterpart, these periods take values in
 \begin{align}\frac{H_2(X_4,\mathbb{R})}{H_2(X_4,\mathbb{Z})}\end{align}
 once we account for identifications provided by large gauge transformations (which have nontrivial cohomology class $[\Lambda_2]\in H_2(X_4,\mathbb{Z})$ ). In this case, the symmetry \eq{2formgs} is just a continuous shift symmetry for the periods of $B_2$.
 
In this Section, we will be interested in a particular solution with nontrivial 2-cycles:  Schwarzschild spacetime. The nontrivial $2$-cycle is the homology class of the event horizon, considering the singularity as excised from the spacetime \footnote{The presence of a non-trivial 2-cycle is actually more manifest in the Euclidean theory, see later.}. Using  the unit sphere $S^2$ volume form $d\Omega$, we can introduce a non-trivial $B$-field background of the form
\begin{align}B=Q\, d\Omega.\label{Bfield}\end{align}
The global symmetry (\ref{2formgs}) with $\Lambda_2=\lambda d\Omega$ shift the $B$-field period $Q$ over the horizon $S^2$ by an amount $\lambda$, and so $Q$ takes values on an $S^1$ of length $1$ (since $\Lambda_2=d\Omega$ corresponds to a large gauge transformation). 

Different values of the $B$ field yield a one-parameter family of  Schwarzschild solutions, all with the same metric independent of the $B$-field value, since $H_3\equiv 0$ and there is no backreaction.

This solution was first discussed by  Bowick, Giddings, Harvey, Horowitz, and Strominger in \cite{Bowick:1988xh}, hence we dub it the BGHHS black hole. The flat  B-field \eq{Bfield} was proposed as an early example of quantum hair on black holes \cite{Coleman:1991ku,Dowker:1991qe}. Classically, the no-hair theorem prevents a four-dimensional black hole from carrying any other information apart from its mass, charge, and angular momentum. The B-field \eq{Bfield}  is however not a local observable; on any contractible region it can be simply gauged away. Therefore, no immediate contradiction arises. Since $H$ vanishes everywhere in spacetime (at least outside of the horizon), turning on this $B$ field does not a priori affect the mass of the black hole or any other property, like e.g. the dynamics of Hawking evaporation. It follows that, at least classically, one can build black holes with arbitrary values of mass and $Q=\int_{S^2}B$. 

If we take the BGHHS black holes as legitimate states, with no further restrictions on the value of $Q$, they render the theory pathological in the same way as remnants do \cite{Susskind:1995da}. Since the symmetry shifts  $Q$, it follows that quantum superpositions 
\begin{align}\ket{n}=\int dQ e^{2\pi inQ}\ket{Q}\label{chargeigenstates}\end{align}
have charge $n$ under the global symmetry. This is a global $U(1)$ charge, which can be added to a black hole with no backreaction on the metric. Notice that this is not the same as the $U(1)$ charge associated to the shift symmetry of the dual axion, which is broken by instanton effects; this will be clear later on in the Euclidean picture. 

The charge $Q$ is an example of a ``vortex charge'', and not an electric or magnetic charge. Electrically charged objects coupled to the $B$ field are strings; magnetically charged objects are instantons. The BGHHS black hole is a defect around which there is a nontrivial flat connection of the 2-form gauge field. This is completely analogous to the description of a vortex in in 2+1 dimensions; in fact, in 2+1 dimensions one can turn a standard Wilson line for the BTZ black hole, yielding a lower-dimensional analog of the BGHHS black hole (see e.g. \cite{Kraus:2006wn} for a discussion when Chern-Simons terms are present).  Vortices for 1-form gauge fields are particles in three dimensions or strings in four; vortices for a 2-form gauge field are particles in four dimensions. 

The crucial difference between a vortex and the other objects (charged under gauge potentials) is that there is no {\em a priori} reason why the vortex charge $Q$ should be quantized. Quantization of electric and magnetic charges stems from Dirac's argument, which does not apply to vortices. Certainly, field theory vortices are often quantized; for instance, in the abelian Higgs model, vortex charge is quantized so that the phase of the Higgs field is single-valued (modulo $2\pi$) at infinity. In the vortices we are considering, however, there is no Higgs field whatsoever: the geometry itself supports a flat connection for the gauge field. 

We have discussed how BGHHS black holes hide a global charge, at least semiclassically.   Before attempting to fix this, we will argue that these objects can generically be produced in the theory, and discuss how quantum effects impact the picture.

\subsection{Production of BGHHS black holes}
\label{sec:production}

Perhaps the most straightforward way to argue for the presence of BGHHS black holes is to consider an euclidean instanton producing a pair of black holes such as the ones considered in \cite{Mann:1995vb,Branoff:1998un}. A spacetime containing two black holes has a nontrivial 3-chain which stretches from one horizon to the other. On the boundary of this 3-chain it is possible to turn on $B$-field, with $H=0$ everywhere, and so without changing the asymptotics. Furthermore if the lagrangian only depends on $B$ via $H$, as the cases we have been considering, then the action will remain unchanged. This $B$ field corresponds precisely to turning a charge of $Q$ and $-Q$ in the two black holes. 

Since the amplitude for producing these black holes is nonvanishing, the states should indeed be part of the theory, for arbitrary real $Q$. Notice that the same argument cannot be made for charges affected by Dirac quantization. For instance, if we consider a $U(1)$ theory, there is an instanton producing a par of oppositely charged magnetic Reissner-Nordstrom black holes \cite{Gibbons:1986cq,Garfinkle:1990eq,Dowker:1993bt,Hawking:1994ii}. It seems one can tune the value of the black hole magnetic charge $q_m$ to any real number. However, when computing the actual contribution of the instanton to the path integral, one has to integrate over quadratic fluctuations around the instanton background. These will include charged particles of charge $q_e$ which, when winding $n$ times around the Dirac string connecting the two black holes, will yield a phase $2\pi q_mq_en$ to the amplitude. This will only be well-defined if Dirac quantization is satisfied. 

We could also try to build a BGHHS black hole directly: Consider an ordinary Schwarzschild black hole and a string coupled electrically to the $B$-field near it. If the string starts to wiggle, it will produce $B$-field radiation, some of which will be absorbed by the black hole. After the string settles down, there will be some $H$ scattered all the way to infinity, while the part that falls to the black hole endows it with $Q$-charge. 

\subsection{Quantum effects and observability of quantum hair}\label{sec:observability}

Reference \cite{Coleman:1991ku} addressed the BGHHS black hole, questioning the observability of $Q$-hair. As discussed above, the value of $Q$ is measured by the holonomy operator $\exp(2\pi i f\int_{S^2} B)$, which can be measured via a Aharonov-Bohm effect in which a string circles the black hole. However, this picture ignores the backreaction of the fields sourced by the string. An infinite string sources a logarithmically divergent $B$ field, 
\begin{align}B=\ln(r) dz\wedge dt\end{align}
where $r$ is the radial distance to the string, which extends along the $z$ axis. The energy density of this field curves space so strongly that it is not possible to have flat asymptotics for a single string. This can be shown explicitly: The above $H$ profile corresponds to an axion of the form $\phi\sim\theta$, so that the integral of the axion profile $\int_{S^1} d\phi=2\pi nf$. As this is a topological quantity, this forces nonvanishing $d\phi$ even far away from the string, and in fact using the stress-energy tensor for the scalar field one finds nonvanishing $T\propto (2\pi nf)^2$ at infinity, which contradicts flat asymptotics.

This problem can be cured by considering closed strings, or configurations with strings and anti-strings together. If the string forms a loop of radius $R$, far away from $r$ the fieldstrengths of both halves of the string add up with opposite signs, so that
\begin{align}\lim_{r\gg R}H\approx\left(\frac{1}{r-R}+\frac{1}{r+R}\right) dr\wedge dz\wedge dt\approx \frac{-2R}{r^2}dr\wedge dz\wedge dt,\end{align}
which is indeed compatible with flat asymptotics. However, near the center of the string, the large $H$ field sourced by the string adds up instead of cancelling out. As a result, the energy of the string scales as $R\log(R)$. It is energetically favorable, for large $R$, to nucleate a concentric string with opposite orientation; as a consequence, the energy of the configuration will scale as $R$ instead. Actually, once quantum effects are taken into account, the situation can become much worse: it is possible for the energy to grow as $R^2$, i.e. the strings are confined. Indeed, it  is known \cite{Polyakov:1976fu} that the dimensionally reduced system of a $U(1)$ gauge field in (2+1) dimensions can show confinement, as shown explicitly by the behavior of the Wilson loop. The naive logarithmic dependence of the electrostatic potential between two charges is corrected to a linear one by quantum effects involving instantons. A slightly more detailed discussion of this effect can be found in Appendix \ref{app:polyakov}. Effectively, this means that the $B$-field is now coupled to an emergent three-form, as we will discuss in Section \ref{sec:gauging}.

While these considerations show that the dynamics of lassoing a black hole with a string are far more complicated that one could have expected, they do not show that such a procedure is impossible in principle. The state of a single macroscopic string might be terribly unstable in practice, but it is a legitimate state of the theory and there will be some (incredibly suppressed) amplitude for it to return to its initial state, times some phase. Thus the proposed ``experiment'' might still possible in principle. 

We could also drop any pretense of relating $Q$ to a Aharonov-Bohm experiment and simply try to understand the spectrum of the gauge-invariant operator $\exp(2\pi i\int B)$ in a black hole background. To do this, it is very convenient to switch to the Euclidean picture. The discussion is particularly clear because the Euclidean Schwarzschild solution
\begin{align}ds^2=\left(1-\frac{2M}{r}\right) d\tau^2+ \frac{dr^2}{1-\frac{2M}{r}}+r^2d\Omega_2^2\end{align}
does not have any horizons, and the $2$-sphere parametrized by the angular coordinates is now noncontractible (the representative of minimum area is at $r=2M$). The topology of the solution is thus $\mathbb{R}^2\times S^2$.  The $\mathbb{R}^2$ is the $(r,\tau)$ plane, where $\tau$ is the usual periodic Euclidean time coordinate. The $S^2$ corresponds to the angular variables, and its volume depends on $r$. The Euclidean BGHHS black hole has a nontrivial period of $B$ over the $S^2$. 

The Euclidean perspective suggests a possible solution to the $Q$ charge problem which is not easily discussed in the Minkowskian perspective: The effects of euclidean strings can a priori break the 2-form global symmetry. This is easier to describe by regarding the systems as an $S^2$ compactification to two dimensions, in which the 2-form symmetry of interest becomes an ordinary shift symmetry for the 2d axion $\phi=\int_{S^2}B_2$.

This is superficially analogous to an ordinary 4d gauge field $A$ compactified on $S^1$ to 3d, leading to an axion
\begin{align}\phi=\int_{S^1}A\end{align}
The higher dimensional theory in principle has a $1$-form global symmetry, which shifts the periods of $A$. This reduces to an ordinary global shift symmetry $\phi\rightarrow\phi+c$ for the axion field. However, as is well-known (see e.g. \cite{Ooguri:1996me,Seiberg:1996ns,ArkaniHamed:2003wu,delaFuente:2014aca}, this symmetry is actually broken by the coupling of electric particles in the higher-dimensional theory to the gauge field. A particle of charge $q$ wrapping the $S^1$ has a contribution
\begin{align}e^{-S+iq\phi}\end{align}
to the path integral, which manifestly breaks the continuous shift symmetry of $\phi$. The effect of these particles can be understood as a 1-loop effect in the effective theory involving the tower of KK modes, and provides a periodic potential for $\phi$ in the lower-dimensional theory, such that the continuous symmetry is spontaneously broken. 

One could think that same effect would break the 2-form global symmetry acting on the periods of a gauge field $B$ wrapped on the $S^2$ of the Euclidean Schwarzschild solution. Indeed, reducing on the $S^2$ yields an effective 2d theory on the $(r,\tau)$ plane, an the 2-form global symmetry becomes an ordinary global shift symmetry for the axion $\phi=\int_{S^2}B_2$.  The desired instantons are Euclidean strings wrapping the $S^2$ of smallest volume, at the origin of the $(r,\tau)$ plane. However, the action of this configuration diverges once we take into account backreaction, as will be discussed in more detail in Section \ref{sec:unbroken}.  

 In this way, the symmetry remain unbroken even after accounting for effects of global strings. Consistency with quantum gravity should exploit other mechanisms to remove the shift symmetry of $\phi$, which will be the subject of Section \ref{sec:break}. Before that, let us make a few further comments on how quantum effects impact the observability of quantum hair. 

Classically, the BGHHS charge corresponds, in the 2-dimensional euclidean perspective, to the vev of $\phi=Q$. Thus, we seem to have a continuous shift symmetry which is spontaneously broken. However, this cannot happen at the quantum level, due to the Coleman-Mermin-Wagner theorem \cite{Mermin:1966fe,Coleman:1973ci}, which guarantees no spontaneous symmetry breaking either in the vacuum or in a thermal state. Since the Euclidean Schwarzschild black hole morally represents a thermal ensemble at Hawking temperature, the theorem applies in this case as well\footnote{Strictly speaking, a Schwarzschild black hole in flat space is not a thermal state, because it evaporates. However, the proof of the Coleman-Mermin-Wagner theorem only depends on the IR of the Euclidean solution, and so it carries over to the Euclidean Schwarzschild solution.}. Thus, the expectation value of any charged operator, such as $\exp(i\phi)$, vanishes. Thus the naive semiclassical $Q$ charge actually vanishes in the quantum theory.

To sum up, the Coleman-Mermin-Wagner theorem makes the global charge completely unobservable from outside of the black hole - but at the expense of an infinite degeneracy of black hole microstates: The $Q$ charge leads to the same kind of trouble as any other global charge in quantum gravity, so it should be dealt with accordingly. 

\section{Breaking generalized global symmetries}\label{sec:break}

We have seen in the previous section how exact generalized global symmetries can lead to troubles in quantum gravity.
Therefore, when faced with a global symmetry in quantum gravity, there are essentially only two options:
 \begin{itemize}
 \item Gauging
  \item Breaking
 \end{itemize}


We start reviewing the first option in which the global symmetry becomes the global part of a gauge symmetry associated to an additional $(p+2)$-form gauge field. However, there are examples in which this is not the case and the global symmetry should therefore be broken.
For the rest of the Section we analyze possible mechanisms of breaking the symmetry, including charged states, exotic stringy effects and explicit breaking by coupling to a $(d-p-2)$-form gauge potential. We arrive at the conclusion that the most generic mechanism which works for any dimension and is present in all known (at least for us) examples of string theory is an elaborated version of the latter, in which the explicit breaking comes from the ubiquitous presence of Chern-Simons terms.

\subsection{Gauging}\label{sec:gauging}
Gauging of the shift symmetry of the periods of a $p$-form potential $B_p$ can be achieved generically by introduction of a $(p+1)$ form potential $C_{p+1}$, and modifying the kinetic term to
\begin{align}\label{KScoupling2}\frac12\vert dB_p-mC_{p+1}\vert^2,\end{align}
where $m$ is the gauge coupling. This system enjoys a gauge invariance
\begin{align}B_p\rightarrow B_p+\Lambda_p,\quad C_{p+1}\rightarrow C_{p+1}+d\Lambda_p,\label{gaugsymKS}\end{align}
which includes shifts of periods of $B_p$. For $p=2$ in four dimensions, the case relevant to the BGHHS black hole of the previous Section, this is the Kaloper-Sorbo lagrangian in its dual formulation \cite{Dvali:2005an,Kaloper:2008fb}. 

As mentioned briefly in Section \ref{sec:rev:GGS}, gauging of the symmetry has dramatic effects for the charged objects coupled to it: A charged $(p-1)$-brane would couple electrically to $B_p$ via a worldvolume coupling
\begin{align}\int_{\mathcal{W}_p} B_p,\end{align}
where $\mathcal{W}_p$ it s the brane $p$-dimensional worldvolume. However, it is clear that this is not invariant under the gauge symmetry \eq{gaugsymKS}. However, if $\mathcal{W}_p=\partial\mathcal{V}_{p+1}$, then the operator
\begin{align} \int_{\mathcal{W}_p} B_p-m\int_{\mathcal{V}_{p+1}} C_{p+1}\end{align}
is indeed gauge invariant. In other words, the $(p-1)$ branes cannot exist on their own; they are always the boundary of a $p$-brane which couples electrically to $C_{p+1}$. 

We can always use the gauge freedom \eq{gaugsymKS} to go to a gauge in which $B_p=0$, which makes $C_{p+1}$ massive. This is a higher-dimensional version of the Stuckelberg mechanism; manifest that by gauging the symmetry we are actually in a Higgsed phase for $C_{p+1}$.  In the 2d case, the above mechanism  corresponds to a Stuckelberg coupling of the axion to a 1-form gauge field via $|d\phi+A_1|^2$. Arbitrary shifts of the axion are now symmetries because they can always be compensated by transformations of the gauge field.

We now describe sketchily how gauging helps with the remnant trouble presented in the previous Section. This is best done in the Euclidean perspective. The Euclidean Schwarzschild metric asymptotes to the flat metric on $S^1\times S^2\times \mathbb{R}$. We may turn on a flat $C_3$ connection at infinity,
\begin{align}C_3=\mu dt\wedge d\Omega.\end{align}
This implies that as we parallel-transport a charged operator such as $\exp(in\int B_2)$ around the time circle, it acquires a factor $\exp(-\beta \mu n )$ as it winds  (here $\beta$ is the asymptotic periodicity of the Euclidean time coordinate). The path integral representation of a chemical potential is obtained by specifying that charged operators should pick up a phase after undergoing parallel transport along the time circle. In other words, this asymptotically flat connection describes a chemical potential. 

In other words, by computing the Euclidean action on the Euclidean Schwarzschild solution with nonzero $\mu$, we actually get $Z(\mu)$, the partition function with a chemical potential for the global symmetry turned on. Because of gauge invariance, it must actually depend on the product $m\mu$. In any case, the inverse Laplace transform of $Z(\mu)$ gives $Z_{n}$, the partition function in the sector of charge $n$. Nontrivial $\mu$ dependence means that the $Z_{n}$ are not all equal; in other words, states with different charge $n$ are not mass-degenerate, avoiding the trouble with remnants. We also see that in the $m\rightarrow0$ limit $Z(\mu)$ should be independent of $\mu$,  and thus we recover infinitely many degenerate states and the usual trouble with remnants.

We should also remark that the particular case of interest here, gauging of a $(d-2)$-form global symmetry, can be subtle, in that the gauging potential $C_{d-1}$ does not have propagating degrees of freedom. Because of this, it can often show up ``for free'', as an emergent field which is not part of the tree-level description of the theory. As an illustration, consider again the $B$-field of the previous Section. The  dual axion $\phi$ is coupled to instantons, and these may or may not generate a potential for $\phi$, as discussed in Appendix \ref{app:polyakov}. If they do, then the strings coupled to the $B$-field confine; equivalently, there is an emergent dynamical three-form $C_3$, which couples to $B$ in the Kaloper-Sorbo fashion via eq.\eqref{KScoupling2}, with a noncanonical kinetic term \cite{Dvali:2005an,Garcia-Valdecasas:2016voz}. 

As a result, in this example the symmetry is gauged by an emergent three-form. In any case, there are plenty of examples in string theory in which the shift symmetry is not gauged - this will happen whenever the instantons do not induce a potential for the dual axion - so we must also explore the second option, breaking the symmetry.

\subsection{Presence of charged states}\label{sec:unbroken} 

The first mechanism one might think of to break the global symmetry is the introduction of electrically charged states. For high enough dimension, these states will explicitly break the symmetry via loop effects. However, the action of these objects diverge for $d\leq p+3$.
For instance, we have seen in the previous Section that, in a 4d theory with a 2-form gauge field $B_2$, the  global symmetry acting on the periods of $B_2$ is unbroken, even after introducing quantum effects from charged strings. In terms of the 2d theory obtained upon compactification, the problem can be traced back to the fact that the  instantons which would break the axion shift symmetry actually turn out to have formally infinite action. This generic feature of low-codimension charged objects means that the two-dimensional current is conserved, $d*j=0$. 

From this perspective, it is clear that this is a generic feature of continuous shift symmetries in two dimensions, and we will have the same trouble in any two-dimensional theory of gravity or compactification of string theory to two dimensions. Therefore, it also applies to global symmetries in high dimension if we impose that the theory should make sense upon compactification. We will now establish this in general, for any $(p+1)$-form global symmetry, and discuss the IR divergence that prevents the existence of the required charged objects. We will also consider AdS and dS asymptotics for the metric, which are clearly interesting.

Consider a $p$-form gauge symmetry with gauge potential $C_{p+1}$ in $d$ dimensions (we will work in the Euclidean picture). We will assume that the theory is lagrangian, and that the leading contribution is the standard canonical kinetic term
\begin{align}\frac12\vert F_{p+2}\vert^2.\label{cankin}\end{align}
We will also allow for interactions with other fields, dilaton couplings, etc. as long as they depend only on the fieldstrengths. As described in Section \ref{sec:rev:GGS}, the above theory has electric $(p+1)$- and magnetic $(d-p-3)$-form generalized global symmetries, which act on periods of $C_{p+1}$ and of the magnetic potential.

For the remainder of this Section we will focus on the electric $(p+1)$-form symmetry. For high enough dimension $d>p+3$, the $(p+1)$-form global symmetry can be generically broken by electrically charged objects, as discussed in Section \ref{sec:rev:GGS}. A $p$-brane which couples electrically to $C_{p+1}$ induces a term in the action
\begin{align}\int_{W_{p+1}} C_{p+1},\end{align}
where $W_{p+1}$ is the $(p+1)$-dimensional worldvolume of the $D$-brane, which explicitly breaks the symmetry.

If $\mathcal{C}_{p+1}$ is homologically trivial, the wrapped branes have finite action and therefore the $(p+1)$-form symmetry is very generically broken. However, if $\mathcal{C}_{p+1}$ represents a nontrivial $p+1$-cycle, the branes that break the symmetry carry nontrivial conserved charges. We will now show that  for low dimensionality ($d\leq p+3$) electrically charged objects fail to break the $(p+1)$-form symmetry, which survives as an exact global symmetry. 

The reason for this is that in low dimension and for the action \eq{cankin}, the electric fields sourced by the branes decay so slowly that the action diverges unless the total charge vanishes. This is is the same kind of behavior that leads to the Coleman-Mermin-Wagner theorem. 

More precisely, consider the theory on a manifold with $d\leq p+3$, which also contains a nontrivial $(p+1)$-cycle. Since we will be discussing an IR effect, let us work on the dimensionally reduced theory, which lives in $d'=d-p-1\leq 2$ dimensions. In what follows we will focus in the $d'=2$ case; the results are similar for $d'=1,0$. The generalized global symmetry now becomes an ordinary 0-form continuous shift symmetry of the two-dimensional axion
\begin{align}\phi\equiv\int_{\mathcal{C}_{p+1}} C_{p+1}.\end{align}
Dimensional reduction of \eq{cankin} gives $\phi$ a canonical kinetic term, provided that the volume of $\mathcal{C}_{p+1}$ is finite. We will assume this from now on, and will comment briefly about alternative situations in Subsection \ref{subsec:fr}. Reducing Gauss' law of the higher-dimensional gauge theory yields 
\begin{align}\int * d\phi=Q,\label{glaw2d}\end{align}
where $Q$ is the net $p$-brane charge wrapped on the $(p+1)$-cycle. In any given configuration in the path integral sum, symmetry breaking effects are proportional to $Q$. In $d>p+3$, that is, when $d'>2$, these terms contribute to the path integral and effectively break the symmetry. In $d\leq p+3$, however, \eq{glaw2d} together with the canonical kinetic term for $\phi$ imply that in any such configuration the action suffers from an IR divergence,
\begin{align}\int \frac12 \vert d\phi\vert^2\sim Q^2\int \frac{dr}{r^{d-p-2}}\rightarrow\infty.\end{align}
As a result, a selection rule $Q=0$ is imposed. Because of this, (part of) the $(p+1)$-form global symmetry remains unbroken. The part that survives corresponds to the 0-form global symmetry arising upon dimensional reduction on the $(p+1)$-cycle.  

The conclusion still holds in presence of extra fields or dilatonic couplings: The allowed configurations should asymptote to the vacuum at large enough $r$, and so the dilaton and any other fields should asymptote to their vacuum expectation values. We should require a finite asymptotic value for the dilaton. The gauge field demanded by \eq{glaw2d} also asymptotes to the vacuum, but so slowly that it cannot avoid a divergence. As mentioned in \cite{Coleman:1991ku}, things are even worse when one couples to gravity: Because the gauge field backreacts on the metric, there are no asymptotically flat $Q=0$ solutions at all. 

\subsubsection{The AdS case}
There is an obvious caveat to these considerations, the reliance on flat asymptotics. For most of our examples this is the case of interest, but later on we will also be interested in the AdS case.  The field sourced by the instanton in global Euclidean $AdS_2$ is 
\begin{align}\phi(r)\sim\ln\left(\frac{1-r/l}{1+r/l}\right)+\phi_0\label{abc}\end{align}
where $l$ is the $AdS$ radius. Unlike its flat space counterpart, for large $r$ the above propagator goes as $1/r$. This means that the IR divergence present in flat space is absent, so it would seem that charged states can break the symmetry. The story is a little more subtle, however. To have a well-defined theory in AdS space, we must specify boundary conditions for the fields. In the present case of a scalar field, we can specify either Neumann, Dirichlet, or mixed boundary conditions \cite{Klebanov:1999tb,Marolf:2006nd}. The latter two explicitly break the global shift symmetry of the axion, which forestalls any discussion of spontaneous symmetry breaking; the electric branes merely generate a contribution to the vacuum energy.

On the other hand Neumann boundary conditions forbid a logarithmic field such as in \eq{abc}. As a result, the selection rule $Q=0$ is again imposed. However, in this case, masslessness forces the dual operator to sit right at the unitarity bound \cite{Marolf:2006nd}, and therefore the field is a noninteracting singleton. If we want it to be interacting, the symmetry has to be explicitly broken. In the context of AdS/CFT, it often happens that the same bulk theory with different boundary conditions gives rise to different CFT's. If we want the scalar to be interacting no matter what boundary conditions we impose, we reach the conclusion that the symmetry must be explicitly broken, much as in the flat space case. This is therefore consistent with the absence of the exact global symmetry in quantum gravity. 

\subsubsection{The dS case}
The other case of interest is of course de Sitter space. Euclidean $dS_2$ is just $S^2$ with its standard metric. The selection rule $Q=0$ now becomes Gauss' law. We thus see that this case is qualitatively similar to Miknowski.

\subsubsection{Final remarks}\label{subsec:fr}
So far we have implicitly considered compactification on manifolds without warping, but this ingredient is actually essential for some discussions; for instance, in the Euclidean Schwarzschild metric the volume of the $S^2$ changes as a function of $r$. The two-dimensional axion decay constant will therefore depend on $r$. Intuitively, the two-dimensional instanton action will remain divergent as long as the coupling does not go to zero at $r\rightarrow\infty$. For instance,  the $B_2$ field in Section \ref{sec:hair} leads to a two-dimensional coupling
\begin{align}\frac{1}{2r^2}\vert d\phi\vert^2,\quad \phi=\int_{S^2}B_2.\end{align}
We see that the coupling diverges as $r\rightarrow$ infinite, so the instanton action is still divergent; in fact, it now diverges as $R^3$, rather than logarithmically (as already noticed in \cite{Coleman:1991ku}). On the other hand, $\varphi$, the zero-mode over $S^2$ of the four-dimensional axion dual to $B_2$, now has a two-dimensional coupling which goes as $1/r^2$. This means that the instanton action is now finite (the IR tail of the field goes as $r^{-1}$, like in the AdS case), and the instantons are allowed to contribute to the path integral, as they do in four dimensions. This illustrates that our results require nonsingular kinetic terms for the 2d axion in the deep IR. Otherwise, it is possible for the instantons we consider to have finite action \cite{Banks:2010zn}. \\

To sum up, we have established that charged objects are not enough to ensure the breaking of generalized global symmetries in quantum gravity. In particular, when compactifying to two dimensions, they fail to completely break the global shift symmetry of the resulting axion scalar, so we need another mechanism to guarantee the absence of exact global symmetries.

The above considerations do not mean that low-dimensional charged objects can never break the symmetry; only that the generic mechanism via instantons available in higher dimensions does not work for $d-p<3$.  For instance, chiral matter can result in Chern-Simons terms under compactification, which can break the 2d axion symmetry. The Chern-Simons term is induced via a closed loop of the chiral fermions, which does not have net electric charge and therefore it does not suffer any IR divergence, as expected due to its topological nature. We will see an example in Section \ref{sec:conj}.   Another possibility is a charged scalar with a potential allowing spontaneous symmetry breaking in three dimensions.

\subsection{Symmetry breaking in a holographic example}\label{sec:holex}
Having stated the problem, we will now provide a concrete solution in a controlled holographic example. Notably, the problem of generalized global symmetries is already present in the canonical example of the AdS/CFT correspondence, type IIB string theory on $AdS_5\times S^5$. The IIB RR potential $C_8$ can be wrapped on the $S^5$ to yield a five-dimensional three-form potential, $C_3$, which leads to a problematic 3-form generalized global symmetry in AdS$_5$. Actually, we can phrase the problem in terms of black hole physics as in Section \ref{sec:BGHHS}. In 5d, black holes have a $S^3$ horizon, around which one can wrap a period of $C_3$, resulting in a  five-dimensional AdS version of the BGHHS black hole. 

Our first step is to determine the boundary conditions obeyed by $C_3$. This can be done by noticing that $C_3$ is the dual potential to the RR axion $C_0$. We know that $D(-1)$ instantons are allowed with the usual boundary conditions \cite{Banks:1998nr}, which means that we should sum up over configurations with different values of 
\begin{align}\int_{\partial AdS} F_4\end{align}
in the path integral, where $F_4=dC_3$. Similarly, configurations with nonvanishing $*F_4=dC_0$ would lead to nonconstant $C_0$ near the boundary, which would conflict with the Dirichlet boundary conditions for this field. As a result, we are forced to conclude that $C_3$ is subject to Neumann boundary conditions, which fix the value of the normal derivative $*F_4$ near the boundary. 

Hence, by the standard AdS/CFT dictionary, $*F_4$ sources a deformation of the CFT,
\begin{align}\int_{\rm CFT} *_{5d}F_4 \wedge \omega_3,\end{align}
and the dual operator $\omega_3$ must be a three-form which is well-defined up to a total derivative. It is natural to take $\omega_3$ as the Chern-Simons three form of $\mathcal{N}=4$ SYM. Since $d\omega_3=\tr F^2$, this can be motivated by rewriting the theta term as
\begin{align}\int_{\rm CFT} C_0 \tr (\,F\wedge F)\sim \int_{CFT} *_{5d}F_4\wedge \omega_3,\end{align}
which indeed shows that the normal derivative of $F_4$ on the boundary is a source for the Chern-Simons three-form.  Finally, this picture is just a higher p-form version of a Neumann boundary condition. 

The would-be continuous shift symmetry that $C_3$ enjoys in the bulk is mapped to constant shifts of the Chern-Simons three-form. 
From this point of view, it is clear that the symmetry is explicitly broken: Any field configuration with non-integer Chern-Simons number on the sphere must have a nonvanishing fieldstrength (configurations with $n_{CS}=1/2$ correspond to sphalerons) and hence nonvanishing energy coming from the Yang-Mills kinetic term. This is the so-called sphaleron potential \cite{Manton:1983nd}. 

In the gravity dual, we would like to consider the behavior of $C_3$ period over the non-contractible $S^3$ in the Euclidean Schwarzschild solution associated to the 5d black hole.  To understand the breaking of the 3-form global symmetry, we only need to provide the bulk description of the sphaleron potential, since the value of $C_3$ is allowed to fluctuate and it will settle at the minimum of this potential. The analysis is most intuitive if we consider the dual pair in the thermal state\footnote{The status of the global symmetry in the vacuum can be carried out in the general analysis of Section \ref{sec:conj}, see section \ref{sec:otherexamples}.} \cite{Witten:1998zw}, in which the boundary SYM lives on $S^3\times S^1$.
 
In the high-temperature phase (large $S^3$ regime), the dominant bulk saddle contains an euclidean Schwarzschild black hole. As before, the two-dimensional axion 
\begin{align}\phi=\int_{S^3} C_3\end{align}
obtained by wrapping $C_3$ on the black hole non-contractible 3-sphere seems to inherit a global 0-form shift symmetry.  However, at finite $N$, we should include other saddles of the Euclidean path integral, such as euclidean AdS. In supergravity, setting up some finite value asymptotic value of $C_3$ on Euclidean $AdS_5$ creates an unphysical singularity in the bulk, which formally has infinite action and should not be included in the path integral. However, this is not the case in string theory, where the singularity is smoothed out. For instance, for $C_3=1/2$, the singularity is replaced by an unstable $D0$-brane; this is the bulk description of the CFT sphaleron, see \cite{Harvey:2000qu,Drukker:2000wx }

In other words, the AdS saddle is manifestly not invariant under shifts of $C_3$, but rather scans the sphaleron potential as we shift $C_3$. Computing the potential explicitly in the bulk requires string field theory, as it is equivalent to computing the open-string tachyon potential for the unstable $D0$ tachyon.

Technically, that solves the problem; the shift symmetry of $C_3$ is broken in the full theory, as it should. The fact that it seems to be there in the Schwarzschild-AdS saddle is just a reflection of the fact that it is restored at temperatures much higher than the height of the sphaleron potential. In the infinite temperature limit, where only Schwarzschild-AdS contributes and we can think of the symmetry as exact, the Coleman-Mermin-Wagner theorem ensures that it remains unbroken in the thermal state; as a result, any charged operators, such as $\exp(2\pi i \int C_3)$ must have vanishing expectation value. We conclude, as we did in Section \ref{sec:observability}, that there is no associated quantum hair for the black hole; it is washed out by the quantum mechanics of the zero mode of $C_3$. 

Naively, one would expect symmetry-breaking effects to become dominant when the temperature drops below the height of the sphaleron potential. However, this is not exactly what happens. The Yang-Mills result \cite{Drukker:2000wx}
\begin{align}M_{Sph}=\frac{3\pi^2}{g_{YM}^2R}=\frac{3\pi^2N}{\lambda R},\end{align}
where $R$ is the radius of the $S^3$ where the dual theory lives, is only valid for small 't Hooft coupling $\lambda$. For the large $\lambda$ limit we are interested in, the height of the sphaleron is given by the mass of the unstable $D0$ brane,
\begin{align}M_{D0}=\frac{\sqrt{2}}{g_s\sqrt{\alpha'}}=\frac{4\sqrt{2}\pi N}{\lambda^{3/4}R}=\frac{4\sqrt{2}\pi \lambda^{1/4}}{g_{YM}^2R}.\end{align}
On the other hand, the Hawking-Page phase transition takes place at a temperature temperature $\sim R^{-1}$. So it seems that we transition to the symmetry-preserving phase at a temperature much lower than the height of the sphaleron barrier. This is due to entropic effects. Because the black hole has so many microstates as compared to the AdS contribution, they start to dominate the partition function at temperatures far below their typical energy. Since the symmetry is approximately restored in the black hole phase, we would expect the black hole mass at the Hawking-Page threshold to be greater than the height of the sphaleron potential. The mass of a black hole of radius $R$ is
\begin{align} M_{BH}R\sim\frac{N\lambda}{g_{YM}^2},\end{align}
and since $N\lambda^{3/4}\rightarrow\infty$, this is indeed the case.

It is also interesting to look at how the symmetry-breaking effects look like from the two-dimensional point of view obtained upon reduction on the $S^3$. The AdS and the AdS-Schwarzschild saddles only differ in the fact that the $S^3$ collapses to a point in the former, but not on the latter. From a two-dimensional perspective, one may regard the collapsed $S^3$ on the AdS saddle as a ``bubble'' of sorts - a region of size $\sim\sqrt{\alpha'}$ in which the two-dimensional low-energy theory fails to provide an accurate description of the physics. A nontrivial $C_3$ gets a contribution to its potential energy from this small bubble, which is only well described by stringy physics. The field outside of this stringy region sees no potential; it doesn't backreact or source the two-dimensional axion field, as an instanton would. We interpret this stringy bubble as a particular avatar of the bubbles in the general setup of Sections \ref{BGHHS-CS}, \ref{sec:otherexamples}, albeit localized at the bottom of the AdS gravitational potential well.

\subsection{Field theory symmetry breaking by topological mass}
\label{sec:fieldth}
The AdS/CFT example shows that in specific circumstances exotic stringy physics comes to the rescue to break the problematic global symmetry. However, this is neither the most general nor the simplest possibility. We now show that the breaking can occur purely within the realm of field theory. In fact, in subsequent sections we will argue that (in suitable avatars) this is the general solution seemingly present in generic string vacua.

Let us focus on the global shift symmetry of a 2d scalar axion. Due to the CMW theorem, the breaking cannot be spontaneous, but due to an explicit term in the Lagrangian, which preserves the axion discrete periodicity. This restricts us to terms of the form
\begin{align}\int \phi \, X_2, \label{x2f}\end{align}
where $X_2$ has quantized periods over a compact spacetime.  

A natural possibility is $X_2=NF_2$, with $N$ and integer and where $F_2=dA_1$ is the field strength of an ordinary two-dimensional gauge field. Lifting to a four-dimensional picture in which the axion arises from a 4d 2-form $B_2$, the coupling \eq{x2f} comes from dimensional reduction of a 4d Stuckelberg $BF$ coupling, rendering the 2-form and the $U(1)$ gauge boson massive. In the four-dimensional theory, the periods of $B$ are $\mathbb{Z}_N$-valued, due to Dirac quantization \cite{Banks:2010zn,BerasaluceGonzalez:2011wy}. More in detail, introducing the dual photon $V$ such that $dV=*F$, the Lagrangian can be completed to
\begin{align}\frac12\vert dV-N B\vert^2.\end{align}
This theory enjoys a gauge invariance $B\rightarrow B+d\lambda_1$, $V\rightarrow V+N\lambda_1$. This implies that the monopole operators must be dressed into gauge invariant combinations of the form
\beqa
e^{i\int_L V_1}\, e^{iN\int_{\Sigma} B_2}
\eeqa
where $\Sigma$ is a surfaces with boundary $\partial\Sigma=L$; namely, monopoles with worldline $L$ must expel $N$ strings with worldsheet on $\Sigma$. Conversely, the worldsheet of $N$ strings coupled electrically to $B_2$ can nucleate a hole bounded by a monopole of $A_1$. This means that when using the string to measure the period of $B_2$ on a non-trivial 2-cycle,  $N$ times $\int B$ must contribute a trivial phase; as a result, $\int B$ is $\mathbb{Z}_N$-valued.

We can also see how the quantization coming from the $BF$ terms plays out directly in two dimensions. Upon reduction to 2d, the integral of $*_{4d}F$ on the compactification manifold is constrained to be an integer $k$ because of Dirac quantization\footnote{In a more direct 2d language, without resorting to a 4d lift, one can obtain the same result from the $\phi F_2$ coupling (\ref{x2f}) by arguing for the quantization of $F_2$ over the 2d spacetime, or of its 2d dual $F_0$, by using Dirac quantization for the corresponding domain walls, see later.}, and the effective potential for the two-dimensional axion $\phi$ is $\vert k- N\phi\vert^2$. We recognize the 2d version of the Kaloper-Sorbo lagrangian \cite{Kaloper:2008fb}. There is a multibranched two-dimensional potential, with $N$ minima at $b=k/N$. The potential explicitly breaks the continuous shift symmetry of $b$.

\subsection{Field theory symmetry breaking by Chern-Simons term} \label{BGHHS-CS}

While the $bF_2$ coupling breaks the continuous shift symmetry, it differs from the holographic example above in one important aspect: The breaking is explicit in every point of spacetime, while in the holographic example the symmetry-breaking effects were localized in a ``bubble'' of stringy size. 

A more general problem is that in many (in fact, most) string compactifications, the global symmetries associated to $p$-form gauge fields are neither gauged nor broken (and made massive) by $BF$ couplings (or their generalizations), and there is no evidence for miraculous exotic stringy physics coming down to their rescue. Fortunately, there is a more flexible version of the symmetry breaking mechanism in the previous section, which is compatible with the massless $p$-form gauge fields encountered in many string vacua. We explain it in the following, and moreover claim in Section \ref{sec:conj} that this is actually the general mechanism in which string theory (and arguably any theory of quantum gravity) breaks generalized global symmetries. The mechanism is based on the existence of cubic Chern-Simons couplings in the effective actions, breaking the symmetry explicitly; even in configurations where the vacuum value of the fields seems to render the Chern-Simons irrelevant, the breaking is manifest  the nucleation of bubbles in which the symmetry is broken.

Let us consider the 2d setup with a 0-form global shift symmetry for an axion scalar $\phi$. The key observation is that there is a slight modification of the breaking shift symmetry via a $bF_2$ coupling, implemented by triple Chern-Simons term instead,
\begin{align}\int G_0 \wedge F_2 \, \phi,\label{csBGHHS}\end{align}
where $G_0$ is a new ingredient, a (nondynamical) $0$-form field strength. This is best regarded as the 2d dual of another gauge field strength  $G_2$. The values of $G_0$ are quantized, and (in accordance with the completeness principle \cite{Banks:2010zn}) the theory now contains membranes under which $G_0$ shifts by an integer amount $N$. The theory thus contains different phases, each having a $N\, \phi\, F_2$ coupling with different values of the coefficient $N$.  Electrically charged particles coupled to $G_0$ act as domain walls separating one phase from another. It is clear that the associated current $j=d\phi$ now has a nonvanishing divergence,
\begin{align}d* j= G_0 F_2.\end{align}

If we are in a phase with $G_0\neq0$, then the above argument applies, and the shift symmetry is broken, exactly as in section \ref{sec:fieldth}. However, if $G_0=0$, it would seem that the Chern-Simons term \eq{csBGHHS} does not have any effect, and that the symmetry is unbroken. However, even in this case, we can construct a finite-action configuration of the equations of motion which manifestly breaks the symmetry.  Consider a finite size region with $G_0=N\neq0$. Within it, the $U(1)$ gauge field is in the Higgs branch, eating $\phi$ via a $N\,\phi F_2$ coupling. This is illustrated in Figure \ref{f1}. Therefore, the only presence of the Chern Simons term, which allows for the possibility of nucleating bubbles with $G_0\neq 0$, is enough to ensure breaking of the global symmetry in the full theory.

\begin{center}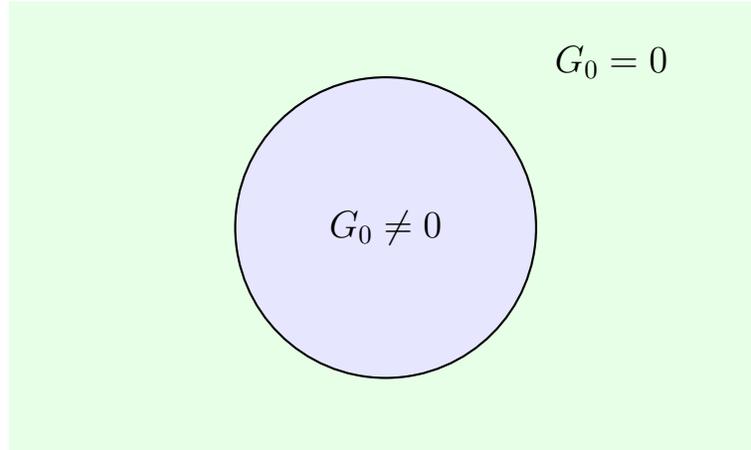
\begin{figure}[!htb]
\centering
\begin{tikzpicture}
\fill[left color = green!01,right color = green!01, middle color=green, bottom color=green!10, top color=green!10](-5,-3) -- (-5,3) -- (5,3) -- (5,-3)-- cycle;
\draw[thick,black,fill=blue!10] (0,0) circle (2cm);
\node at (0,0) {\large $G_0\neq 0$};
\node at (3,2.2) {\large $G_0= 0$};
\end{tikzpicture}
\caption{Triple Chern-Simons terms mean that we must include into the path integral configurations with a bubble in which $G_0\neq0$. Outside of the bubble, in the green area, $\phi\sim\phi+c$ leaves the action invariant, but in the blue area it does not. As a result, $\phi\rightarrow\phi+c$ is no longer a symmetry.}\label{f1}
\end{figure}\end{center}

We can take a step back and see how the triple Chern-Simons terms affect e.g. the BGHHS black hole of Section \ref{sec:BGHHS}. We now have a triple $G_0 B_2 F_2$ Chern-Simons term in the action, and can nucleate bubbles within which the gauge field is Higgsed. Within the bubbles, the value of the $B$-field is quantized, and related to the electric charge of the black hole modulo $G_0$ \cite{Coleman:1991ku}; even if $B$ was freely fluctuating outside, it falls to a definite value within the bubble. This is a reflection of the Kaloper-Sorbo potential that the two-dimensional axion feels within the bubbles. 
We thus break the symmetry via localized configurations which however do not lead to a IR divergent action, as instantons do. This is similar to the stringy bubble in the holographic example, with the advantage that there is a simple field theory description both within and outside of the bubble. A second difference is that the stringy bubble in the holographic setup is an actual (euclidean) saddle which provides contributions to the potential; whereas here
we have not claimed the field-theoretical bubbles to be actual saddles - we have merely shown that they generically contribute to the path integral, and thus break the symmetry. 

Finally, as we will see in Section \ref{sec:examples}, it is also possible to have bubbles with an effective $N \phi F_2$ term inside the bubble, even in vacua when the axion symmetry has been gauged by a second $U(1)$ $A'$, and the lagrangian is $\vert d \phi - p A'\vert^2$.  In this case, the gauge symmetry of $A'$ only shifts $\phi$ outside of the bubble - and is an exact symmetry of the theory. But the global symmetry, which shifts $\phi$ by a constant everywhere, is indeed broken, as above. 

\medskip

Many classes of string compactifications providing explicit examples of the Chern-Simons mechanism are presented in Section \ref{sec:examples}, suggesting this is the general solution realized in string theory, and arguably in any theory of quantum gravity. Accordingly, we draw some general lessons in the next subsections.

\section{A conjecture about Chern-Simons terms}\label{sec:conj}
We have now formally solved the problem stated in Section \ref{sec:hair}: We understand how 2-form global symmetries in four dimensions can lead to a remnant problem, which is not solved by symmetry-breaking effects from charged objects (namely, there are no finite-action instantons in the two dimensional version), but we have learnt the mechanism to break the symmetries explicitly.

As detailed in the next Section, we have found that the triple Chern-Simons terms \eq{csBGHHS} which give rise to field-theoretic symmetry breaking bubbles are present in all the examples we have considered. For the remainder of this one, we will upgrade our observations to a rule, and will propose that consistent theories of quantum gravity suffer from a Chern-Simons pandemic, i.e. \emph{every consistent theory with higher $p$-form potentials and weakly coupled gravity there should have the appropriate Chern-Simons terms}, so that no global symmetries actually survive in two dimensions no matter how we compactify. Equivalently, for any global symmetry in string theory, it seems that we can always find phases in which the symmetry is either gauged or explicitly broken, and there are always domain walls taking us from one to the other. Our conjecture elevates this generic property to a requirement for any consistent theory. 

It is worth emphasizing that we are requiring a consistent quantum theory of gravity to be well-behaved under  compactification. Such a requirement is present in several recently proposed Swampland conjectures, such as \cite{Heidenreich:2015nta,Heidenreich:2016aqi} or \cite{Ooguri:2016pdq}.

This conjecture, if correct, is another example of a Swampland criterion \cite{Vafa:2005ui}, which allows us to further discriminate between effective field theories with and without a sensible UV completion. If correct, it strongly suggests that the Chern-Simons terms ubiquitous in string theory are a generic feature of a consistent theory of quantum gravity. Like any other conjecture, the real support for it comes from examples: In all the stringy examples we have been able to come up with, the conjecture seems to hold. Even the original holographic example discussed in Section \ref{sec:holex} admits field-theoretic bubbles related to Chern-Simons terms, as we will discuss in section \ref{sec:otherexamples}.

In the following we will display the rationale of the conjecture and the arguments why the Chern-Simons provide in our view the most generic mechanism for breaking the global symmetries. Then we will discuss some examples of effective theories which do not fulfill our conjecture as well as possible implications for the SM, leaving the exposure of examples in string theory supporting the conjecture for Section \ref{sec:examples}.

\subsection{Rationale for the conjecture}\label{sec:why}
The examples in the next Section provide evidence for the conjecture in a variety of setups, where the Chern-Simons terms break the generalized global symmetries very generically. Still, this can only be part of the answer, because as illustrated in Section \ref{sec:holex}, two-dimensional symmetry breaking can also be accomplished by stringy effects without a simple 2d field theory description. From a four-dimensional point of view, however, what happens in Section \ref{sec:holex} is clear: The global 3-form symmetry needs a nontrivial bulk 3-cycle to exist and, because gravity forces us to sum over topologies with the same asymptotics, we must include configurations where the boundary 3-cycle is contractible and there is no 3-form symmetry at all.

We see no reason however to think that this mechanism will work generically for any low-dimensional compactification, especially in the Minkowski case. Take e.g. Euclidean quantum gravity on $\mathbb{R}^2\times T^2$, plus a gauge field (but no charged particles). Consider one of the global 1-form symmetry for the periods of $A$; a Taub-Nut background seems to break the symmetry because the $S^1$ is contractible. However, the Taub-Nut is a KK monopole, and because one of the transverse directions is compact, it has real codimension 2 in the non-compact space, so its action has the same kind of IR divergence as discussed in Section \ref{sec:unbroken}. If we were compactifying to $AdS_2\times T^2$ instead, whether or not the KK monopole is allowed would depend on the boundary conditions for the KK $U(1)$.  By contrast, the triple Chern-Simons term discussed in Section \ref{sec:break} break the symmetry in a very generic way. In fact, in a certain sense, the Chern-Simons terms allow, in certain cases, instanton-anti instanton pairs to break the symmetry, giving the two-dimensional case some resemblance to its higher-dimensional siblings. We elaborate on this point in Appendix \ref{app:fat}.

The Chern-Simons terms may be understood from a different point of view in the context of the AdS/CFT correspondence. Consider a bulk AdS  theory with a $p$-form gauge potential $C_p$. It is part of the standard dictionary \cite{Klebanov:1999tb} that different boundary conditions for the bulk fields result in different CFT's. In the example of Section \ref{sec:holex}, the appropriate boundary conditions demand that the bulk $p$- form gauge field be dual to a $p$-form boundary field (the boundary Chern-Simons form), rather than a conserved $p$-form boundary current. 

It is not true in general that one can pick whatever boundary conditions and end up with a consistent theory. Nevertheless, \emph{if} it is possible to have a consistent theory in which the bulk $p$-form gauge field is dual to a boundary current $j_p$, the bulk Chern-Simons terms ensure that $j_p$ is not conserved. This is familiar from the $p=0$ case, where Chern-Simons terms provide the bulk description of an anomalous current \cite{Witten:1998qj}. Bulk Chern-Simons terms are the only way to break invariance under large gauge transformations via bulk terms without also spoiling the local gauge invariance of $C_{p}$.

Thus, in this picture, bulk Chern-Simons terms merely represent the absence of continuous higher $p$-form symmetries in the dual field theory\footnote{This aligns nicely with the fact that non-abelian gauge theories do not admit continuous electric/magnetic $1$-form global symmetries, since the center of the group is discrete \cite{Gaiotto:2014kfa}: Since gauge strings are usually dual to fundamental strings in the bulk description \cite{Maldacena:1997re}, this means that there should be a Chern-Simons term involving $B_2$. Indeed, in the standard $AdS_5/CFT_4$ example, we find such a term $\int N B_2 \wedge dC_2$, which breaks the global part of the $B$-field $U(1)$ 1-form symmetry to $\mathbb{Z}_N$. }. For the case of a $p$-form field in $AdS_{p+2}$ (this includes the $C_3$ field discussed in Section \ref{sec:holex}), we can make a more compelling argument. If it is consistent to choose boundary conditions such that the dual operator is a boundary current $j_p$, compactification to three dimensions should give a gauge field in $AdS_3$, dual to a boundary current. Conformal invariance demands this current to be anomalous, which translates to a Chern-Simons term involving $A$ the bulk theory \cite{Kraus:2006wn}\footnote{These comments are very sketchy - the current can in principle mix with other currents, or compactification may change the sign of the vacuum energy so that there is no $AdS_3$ solution. We merely want to emphasize that the Chern-Simons terms we use to break two-dimensional symmetries can under some assumptions be related to familiar CFT phenomena. }.

To sum up, it is unclear that the gravitational effects that work in Section \ref{sec:holex} can break the symmetry in generic situations, both in Minkowski and in AdS examples with the ``wrong'' boundary conditions. On the other hand, the Chern-Simons terms induce a generic breaking of the symmetry, and are required by the AdS/CFT correspondence in some cases.

\subsection{Implications for Einstein gravity and supergravity}
\label{sec:implic}

To illustrate the power of our Chern-Simons Swampland criterion, we discuss some interesting theories which do not comply with it, and which we therefore claim are not compatible with quantum gravity.

\medskip

$\bullet $ {\bf Einstein gravity in $d\geq 4$}

As a first simple example, pure gravity in $d\geq4$ dimensions is incompatible with our conjecture\footnote{This would imply that the asymptotic safety program \cite{Weinberg:1976xy,Niedermaier:2006ns} should not work, at least without inclusion of matter fields.}. This is easily seen by compactifying four-dimensional gravity on a $T^{2}$: We get a unstabilized 2d axion $\phi_{\rm KK}$, the real part of the torus complex structure. Performing the compactification in two steps, there is a 3d KK $U(1)$ gauge boson $A_{\rm KK}$ in the 4d $\to$ 3d compactification, and $\phi_{\rm KK}$ corresponds to its Wilson line in the 3d$\to $ 2d $S^1$ compactification. The 0-form global shift symmetry for this scalar requires some Chern-Simons term for its breaking, which is not present in the theory, so it is inconsistent. We can reach the same conclusion in the case $d>4$ by compactifying to four dimensions on a manifold without isometries  and then proceeding as before \footnote{For $d=3$ pure gravity is also incompatible with a strengthening of our conjecture which we discuss in Section \ref{sec:3forms}}. 

A simple modification of the theory to render it consistent with our criterion is to e.g. include a 4d axion $\varphi$. This allows to consider compactifications with a Scherk-Schwarz ansatz \cite{Scherk:1979zr}, as follows: A general axion $\varphi\sim\varphi+2\pi$ which is compactified on a circle with periodic coordinate $x\sim x+2\pi$ admits a boundary condition
\begin{align}\varphi(x+2\pi)=\varphi(x)+2\pi n.\end{align}
This means that $e^{i\varphi}$ has charge $n$ under the KK photon, and thus, the gauge-invariant quantity in the dimensionally reduced theory is $d\varphi- nA_{\rm KK}$. This provides a mass for the KK photon, implying that, after further compactification to two dimensions, the axion $\phi_{\rm KK}$ obtained from its Wilson line is also massive and the global shift symmetry is broken. 

This is equivalent to the Chern-Simons criterion, as follows:  in terms of $\eta$, the 2d axion dual to the axion $\phi$, there is triple Chern-Simons $n\,\eta \,F_{\rm KK}$ term, where $n$ should be regarded as a background geometric flux. According to our general discussion, even in the vacuum with $n=0$ there are effects from bubbles nucleating regions with $n\neq 0$ which suffice to break the global symmetry.

This solution, more than a particular case, turns out to be generic in string compactifications, where gravity is always accompanied by the antisymmetric two-form field $B_2$. Upon T-duality, the Stuckelberg coupling  $d\phi- nA_{\rm KK}$ transforms into a Chern-Simons coupling involving $B_2$ and the dual field to the axion, as we will explain in section \ref{sec:examples}.

\bigskip

$\bullet $ {\bf Einstein-Maxwell theory}

Similarly, the conjecture puts four-dimensional Einstein-Maxwell theory in the Swampland. Of course, we knew as much from the WGC \cite{ArkaniHamed:2006dz}: Einstein-Maxwell must be coupled to light superextremal states with $m\leq g M_Pq$.  However, we claim that even this version of Einstein-Maxwell with charged matter is in the Swampland, since upon reduction on a $T^2$ the theory lacks one of the required triple Chern-Simons terms: Upon reduction on the first $S^1$, one obtains an axion $\phi_{A_1}$ (where the subscript reminds us that this axion comes from the gauge field), two gauge fields $A, A_{KK_1}$ (the latter being a KK photon), and the metric. Reducing again, we get three axions in two dimensions: $\phi_{A_1},\phi_{A_2}$ coming from holonomies of the gauge field, and $\phi_{KK}$, coming from reduction of $A_{KK_1}$. The axion $\phi_{KK}$ may be identified with the real part of the complex structure parameter of the $T^2$ discussed in the previous paragraph.  We also have three gauge fields, $A,A_{KK_1}$, and $A_{KK_2}$.

The axion $\phi_{A_1}$ can obtain a potential if we take a Scherk-Schwarz ansatz. In this way, we can obtain a Chern-Simons term for (the dual of) $\phi_{A_1}$ with $A_{KK_2}$, but not for $\phi_{A_2}$ or $\phi_{KK}$. In this  case, the theory can be made compliant with the conjecture above by introducing 4d charged chiral matter, which gives rise to three-dimensional Chern-Simons terms via the parity anomaly both for $A$ and $A_{KK_1}$. Upon dimensional reduction, the 3d Chern-Simons term gives the desired 2d terms for $\phi_{A_2}$ and $\phi_{KK}$.  The trouble with $\phi_{A_2}$ can also be solved by including Chern-Simons terms for the gauge boson $A$ already in four dimensions. 

$\bullet $ {\bf Four dimensional $\mathcal{N}=8$ Supergravity}

A final notable example of theory which does not comply with our criterion, and we therefore claim belongs to the Swampland, is 4d $\mathcal{N}=8$ Supergravity. Simply put, the theory contains a 2-form gauge fields leading to a 2-form generalized global symmetry without the Chern-Simons terms required to break it. Actually, this theory has already been claimed to be in the Swampland, by different arguments related to the impossibility of reaching is as a suitable decoupling limit of toroidally compactified string theory \cite{Green:2007zzb}. In our case, consistency of the theory could be achieved by including the necessary Chern-Simons terms, in particular those involving the Romans mass parameter (and other in its U-duality orbit), which are present in string models as required by our conjecture.

There is a line of work (see e.g. \cite{Bern:2006kd,Beisert:2010jx,Kallosh:2010kk}) trying to figure out whether or not $\mathcal{N}=8$ SUGRA is perturbatively finite. Our statement does not directly exclude this  potential result, but renders it less relevant. Even if the theory is perturbatively finite, it misses the non-perturbative domain walls allowing to interpolate to vacua where the generalized global symmetries are broken, so the corresponding black hole remnant problems excludes it as a complete theory of quantum gravity.

\section{Examples} \label{sec:examples}
We now discuss several classes of examples which comprise the main evidence for our conjecture.

\subsection{KK photons}
One prominent example of $U(1)$'s in effective field theories arising from compactification is that of KK photons, arising from continuous isometries of the internal manifold. These gauge bosons lead to 1-form generalized global symmetries, and in general lack the topological mass or Chern-Simons couplings ensuring their breaking. These theories in general fall in the Swampland, as earlier discussed examples.

However, in string theory these KK photons always turn out to have the required set of Chern-Simons couplings. This can easily shown in KK compactification on $S^1$, leading to a $U(1)$ KK gauge boson; for concreteness we focus on type II theories and a KK photon in a 5d $\to$ 4d compactification, other examples being similar. In order to see the Chern-Simons couplings, let us perform a T-duality along the orbit of the $U(1)$ isometry in the compact space, in which the KK photon turns into the dimensional reduction of the T-dual NSNS $B$ field. The 10-dimensional action has the standard Chern-Simons terms involving the $B$ field,
\begin{align}\int B \wedge F_p\wedge F_{d-p-2}\end{align}
which upon dimensional reduction provide Chern-Simons terms involving $A$ in the four-dimensional effective field theory. In the original frame, the Chern-Simons term (in the form of a Stuckelberg Lagrangian) arises from the Scherk-Schwarz compactification of some RR field. 

As a concrete example, consider the  compactification of type IIA on $M_5\times S^1$. Compactifying on $M_5$ first, we get an axion field $\phi$ coming from the period of $C_5$ on $M_5$. Further compactifying on $S^1$ using a Scherck-Schwartz ansatz $\phi(y+2\pi R)=\phi(y)+2\pi N$ translates to a Stuckelberg lagrangian in four dimensions. The coupling 
\begin{align} N d\phi \wedge * A\end{align}
can be integrating by parts,into $BF$ coupling between the KK photon $A$ and the 2-form $C_2$ 4d dual to $\phi$. In the T-dual frame, such a coupling arises from dimensional reduction on the circle of the Chern-Simons coupling $B_2 \wedge F_3\wedge F_5$, leading to $G_0B_2\wedge F_3$ in five dimensions, with $G_0=\int_{M_5}F_5$. This $F_5$-flux corresponds precisely under T-duality to the axionic flux $N=\int_{S^1}d\phi$ arising from the Scherk-Schwartz reduction.

This class of examples underlie the solution to the problems with pure Einstein theory mentioned in Section \ref{sec:implic}.

\subsection{Type IIA/B flux compactifications}
\label{sec:fluxed}

Probably the richer class of examples supporting the conjecture about Chern-Simons couplings are the four-dimensional effective theories obtained from flux compactifications of type IIA/B string theory. In short, the CS couplings in 10d ensure that all generalized global symmetries coming from higher RR or NS p-form fields are broken or gauged\footnote{It can happen that the CS coupling does not appear in the 10d supergravity action since involves the presence of locally non-geometric fluxes. But they will appear as geometric CS terms in the T-dual theory, as we explain in section \ref{sec:otherexamples}}. 

For concreteness, let us consider again a 2-form field $B_2$ with a 2-form global symmetry in four dimensions. The couplings inducing the breaking of the symmetry are of the form $G_0 B_2 F_2$ as explained in section \ref{BGHHS-CS}, while couplings inducing its gauging occurs arise from higher dimensional version of a Stuckelberg coupling $G_0' (*_{4d}B_2)F_4$ as explained in section \ref{sec:gauging}. Notice that both couplings can arise from higher dimensional CS terms involving either $B_2$ or its dual field. In terms of the axion dual to $B_2$, namely $d\phi=*_{4d}dB_2$, the above couplings are also responsible for gauging or breaking the 0-form global symmetry, but yielding the opposite result. The term  $G_0 B_2 F_2$ breaks the 2-form global symmetry while gauges the 0-form global symmetry, and $G_0' \phi F_4$ gauges the 2-form global symmetry while breaks the 0-form global symmetry. 

It is an important point that both kinds of 4d terms, Stuckelberg and Kaloper-Sorbo, cannot be present at the same time in the effective theory, since their simultaneous presence would lead to a Green-Schwarz anomaly, as mentioned in footnote \ref{foot-gs}. When the couplings arise both from bulk couplings involving just fluxes, their simultaneous presence should in principle be avoided by the consistency conditions on fluxes, like the quadratic constraint in the embedding tensor description of fluxes, see e.g. \cite{Samtleben:2008pe} for a review. 

The general lesson is that, given a 4d 2-form gauge field, the CS couplings are such that there exist domain walls interpolating to phases with Stuckelberg $BF$ couplings, and domain walls to phases with Kaloper-Sorbo couplings, yet there cannot be phases with both couplings present. This implies that in compactifications where the internal fluxes induce a 4d Kaloper-Sorbo coupling which gauges the global symmetry, there are no $BF$ terms breaking it. On the other hand, the theory consistently incorporates domain wall bubbles inside which the flux inducing the Kaloper-Sorbo coupling is absent, but it is precisely in the absence of this coupling that phases with a Stuckelberg coupling can be turned on. 

One may worry about the fate of the gauge symmetry in phases with KS couplings gauging the global symmetry, once we nucleate a bubble in whose interior the symmetry is broken by $BF$ couplings. The above picture is happily self-consistent, since  bubbles with $BF$ couplings necessarily have the KS coupling turned off. Therefore the gauge symmetry only acts on the exterior KS phase and is unbroken; the global symmetry however, which acts both on the KS and BF phases is broken in the presence of the bubble.

 In the following we present a simple example which suffices to illustrate this general pattern of Kaloper-Sorbo vs Stuckelberg couplings.
Let us consider Type IIA compactified on a Calabi-Yau three-fold (without orientifolds), which for simplicity we take with one 3-cycle $A$ and its dual $B$. We define
\beqa
\int_A H_3=p, \quad\int_A C_5=b_2',\quad
\int_A C_3=\phi,\\
 \int_B H_3=p',\quad\int_B C_5=b_2,\quad\int_B C_3=\phi'
\eeqa
where $*d\phi=b_2$ and $*d\phi' = - b_2'$. By dimensionally reducing the 10d CS couplings
$H_3 C_3 F_4$ and  $H_3 C_5 F_2$ we get in four dimensions
\beq
\mathcal{L}\sim (p \phi' - p' \phi) F_4 + (p b_2 - p' b_2' ) F_2
\eeq
The 2-form dual to $(p \phi' - p' \phi) $ is $(p b_2' +  p' b_2)$. Therefore, the two terms in the above Lagrangian actually correspond to two orthogonal 2-form fields. For the first one the 2-form global symmetry is gauged while for the second one is explicitly broken. Notice however that we can always nucleate a domain wall that changes the flux values to $q=-p', q'=p$, in such a way that both scalars interchange roles. The global symmetry that was gauged is now broken, and viceversa. In other words, for a fixed combination of fields, when the coefficient of the Kaloper-Sorbo coupling vanishes and the global symmetry is not gauged it becomes possible to turn on  phases with $BF$ couplings, which break the symmetry.

To recap, as we explained in section \ref{BGHHS-CS}, the presence of the CS term $G_0b_2F_2$ is enough to break the global symmetry, even in the phase with $G_0=0$. However, the same does not apply for the term gauging the symmetry. In that case, if $G_0=0$ asymptotically, we recover an exact global symmetry. We have seen, though, that we can always nucleate a bubble in which the coupling responsible for gauging the symmetry disappears and is replaced by a term explicitly breaking the symmetry. The presence of some regions in space-time in which the global symmetry is broken is enough to guarantee that the symmetry is broken in the full theory. Therefore, in the previous example, all the global continuous symmetries are indeed broken in the full theory.

\medskip

One may fear that the above situation is non-generic, and that the presence of orientifold projections may remove some of the required couplings. We will now illustrate that the details of the discussion change, but the general conclusions remain valid.
Let us consider the addition of O6-plane, and consider $A$ to be odd under the $\IZ_2$ involution. The orientifold projection only allows for the combinations $\int_A H_3=p, \int_B C_3=\phi', \int_A C_5=b_2'$. The only 4d coupling surviving the orientifold projection is $p\phi' F_4$. Now we cannot play the same game as before to interchange the couplings, so one may worry about the phase with $p=0$, in which the symmetry is not gauged. However, if we turn off the flux by setting $p=0$, it is now possible to add D6-branes wrapping a 3-cycle $\Pi$ (and their orientifold images D6' wrapping $\Pi'$), satisfying $[\Pi]-[\Pi']=[A]$, which were not possible for $p\neq 0$. From the brane worldvolume we get a new coupling  $\int_{D6} C_5 \tr F_2$ which leads upon dimensional reduction to $b_2' \tr F_2$ in four dimensions. The fact that the D6-branes are only allowed to wrap a 3-cycle with component along $A$ when $p=0$ is due to a Freed-Witten anomaly induced on the brane when the flux is not vanishing. There is indeed a domain wall interpolating between the configuration with flux (and branes wrapping 3-cycles with vanishing component along $A$) and the configuration without flux (and branes wrapping the above $[\Pi]$ 3-cycles). This domain wall corresponds to a 4-chain connecting the 3-cycles wrapped by both sets of branes\footnote{The existence of these interpolating domain walls introducing wrapped D-branes allows to regard the $BF$such couplings as yet another avatar of cubic Chern-Simons couplings. This is particularly explicit in models like magnetized D-branes, in which certain homological charge classes are associated to world-volume magnetic fluxes on the D-branes, see \cite{Ibanez:2012zz} for review.}. As we cross the four-dimensional domain wall, the brane configuration changes as illustrated in Figure \ref{f2}. The second configuration in the Figure can also be understood in terms of the field theory of the $O(2)$ gauge theory of a $D6$-brane sitting on top of the orientifold, as the soliton with nontrivial charge under $\pi_0(O(2))=\mathbb{Z}_2$. 

\begin{center}\begin{figure}[!htb]\centering
\begin{subfigure}[t]{0.28\textwidth}
\centering\resizebox{.95\textwidth}{!}{\begin{tikzpicture}
\draw[thick] (-3,-3) -- (-3,3) -- (3,3) -- (3,-3)-- cycle;
\draw[thick] (0,3) -- (0,-3);
\draw[dashed] (-3,3) -- (-0.75,0.75);
\draw[dashed] (-3,-3) -- (-0.75,-0.75);
\draw[dashed] (-0.75,0.75) arc (45:0:1.07cm);
\draw[dashed] (-0.75,-0.75) arc (-45:0:1.07cm);
\draw (3,3) -- (0.75,0.75);
\draw (3,-3) -- (0.75,-0.75);
\draw (0.75,0.75) arc (135:180:1.07cm);
\draw (0.75,-0.75) arc (-135:-180:1.07cm);
\node at (0,-3.5) {\huge $A$};
\node at (3.5,0) {\huge  $B$};
\node at (0,3.5) {\huge  $O6$};
\node at (2.3,1.25) {\huge  $D6$};
\node at (-2.3,1.25) {\huge  $D6$'};
\end{tikzpicture}}\end{subfigure}
\begin{subfigure}[t]{0.04\textwidth}\centering\vspace{-2.3cm}$\rightarrow$\end{subfigure}
\begin{subfigure}[t]{0.28\textwidth}
\centering  \resizebox{.95\textwidth}{!}{\begin{tikzpicture}
\draw[thick] (-3,-3) -- (-3,3) -- (3,3) -- (3,-3)-- cycle;
\draw[thick] (0,3) -- (0,-3);
\draw[dashed] (-3,3) -- (0,0);
\draw[dashed] (-3,-3) -- (0,0);
\draw (3,3) -- (0,0);
\draw (3,-3) -- (0,0);
\node at (0,-3.5) {\huge $A$};
\node at (3.5,0) {\huge  $B$};
\node at (0,3.5) {\huge  $O6$};
\node at (2.3,1.25) {\huge  $D6$};
\node at (-2.3,1.25) {\huge  $D6$'};
\end{tikzpicture}}\end{subfigure}
\begin{subfigure}[t]{0.04\textwidth}\centering\vspace{-2.3cm}$\rightarrow$\end{subfigure}
\begin{subfigure}[t]{0.28\textwidth}
\centering
\resizebox{.95\textwidth}{!}{\begin{tikzpicture}
\draw[thick] (-3,-3) -- (-3,3) -- (3,3) -- (3,-3)-- cycle;
\draw[thick] (0,3) -- (0,-3);
\draw[dashed] (-3,3) -- (0,0);
\draw (-3,-3) -- (0,0);\draw (3,3) -- (0,0);
\draw[dashed] (3,-3) -- (0,0);
\node at (0,-3.5) {\huge $A$};
\node at (3.5,0) {\huge  $B$};
\node at (0,3.5) {\huge  $O6$};
\node at (2.3,1.25) {\huge  $D6$};
\node at (-2.3,1.25) {\huge  $D6$'};
\end{tikzpicture}}\end{subfigure}
\caption{Schematic depiction of the transition that takes place across the domain wall described in the main text. As we cross the domain wall, the $D6$ brane approaches the $O6$ plane and reconnects differently with its orientifold image. In the first picture, the $BF$ couplings of the $D6$ brane and its image cancel; in the third, they add up.}\label{f2}
\end{figure}
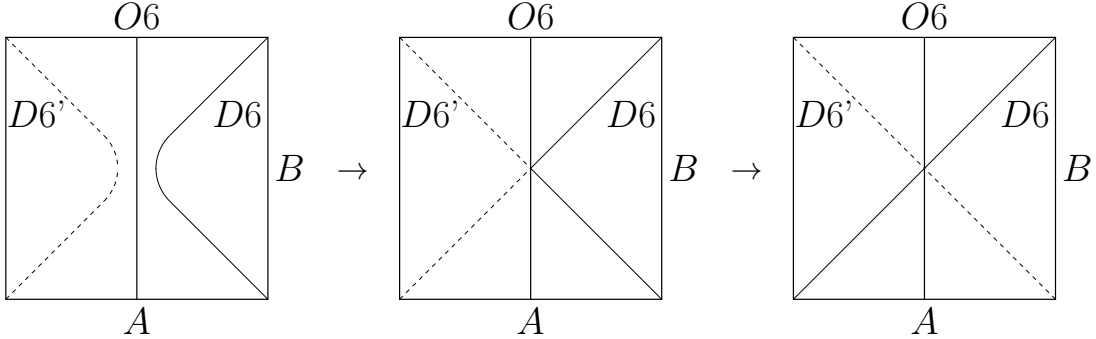  \end{center}

 Therefore, starting with a vacuum in which the scalar field coming from $C_3$ has a $p\phi F_4$ coupling, we can nucleate a bubble with $p=0$ and with the scalar field enjoying a Stuckelberg coupling instead (a $b_2F_2$ for the dual 2-form). The presence of this coupling in some region of the space-time is enough to ensure that the global symmetry is broken in the full theory.

\subsection{Digression: An M theory BGHHS black hole}

The very explicit realization of the Chern-Simons mechanism in terms of fluxes allows the following explicit discussion of (close cousins) of the BGHHS black hole in M theory. In the M-theory  configuration we consider, the BGHHS black hole becomes a system of branes in a horizonless geometry, which will allow us to understand how the associated problems are resolved from an ultraviolet point of view.

Consider compactification of type IIA on e.g. a six-torus  to four dimensions, and a stack of $N$ coincident D6-branes wrapped on the compact manifold. The M-theory lift of the D6 brane stack is a multi-center Taub-NUT metric \cite{Sen:1997js} (see also \cite{Witten:2009xu})
\begin{align} ds^2=U(d\vec{x}\cdot d\vec{x})+U^{-1}(d\theta + M),\quad U=\frac{N}{r}+\frac{1}{\lambda^2}.\end{align}
Here, the Taub-NUT space is described as an $S^1$ fibration over $\mathbb{R}^3$, with $\theta$ being the coordinate along the circle, while $\vec{x}$ labels points in $\mathbb{R}^3$. Finally, $M$ satisfies $dM=*dU$. This spacetime has $N-1$ vanishing 2-cycles, which translate to $N-1$ normalizable Poincare dual 2-forms, and an extra non-normalizable self-dual 2-form $\omega$ which corresponds to the center-of-mass motion of the system. An explicit expression for $\omega$ in the case $N=1$ can be found in \cite{Witten:2009xu}. For general $N$, the center-of-mass self-dual 2-form is given by (see Appendix \ref{app:TNdetails})
\begin{align}\omega=d\Lambda, \Lambda=F(r)(d\theta+M),\quad F(r)=\frac{r}{r+N\lambda^2},\label{selfdualmaintext}\end{align}

Now, consider turning on M-theory 4-form flux $G_4$ along $\omega\wedge \omega$. A valid $C_3$ potential for this is 
\begin{align}C_3=Q\omega\wedge\Lambda=-QNFd\Omega \wedge (d\theta+M).\end{align}
At large $r$, the coordinate $\theta$ parametrizes the $M$ theory circle, and one can decompose $C_3$ in terms of $C_3^{IIA}$ and $B_2$ components, obtaining
 \begin{align}C_3^{IIA}=-QN MFd\Omega\quad B= -QNFd\Omega.\end{align}
We see that at large $r$, the asymptotics is $B\rightarrow - QNd\Omega$, so that we have obtained a microscopic description of a (charged, extremal, singular cousin of a) BGHHS black hole. Although the cost of turning on $Q$ is infinite in type IIA string theory because of the singular $H$ field at the origin, the uplift to M-theory smoothes out the resulting profile for the $B$-field. The $B$ field at the origin, $1/(N\lambda^2)$, is controlled by the asymptotic radius of the Taub-NUT, that is, by the IIA coupling. We see that it indeed diverges in IIA.

It might seem that one can turn on any $G_4$ flux, and hence, that the charge $Q$ is not quantized. To understand what is going on in detail, we need to solve the supergravity equations of motion; the flux $G_4=Q\, \omega \wedge \omega$ will not be a solution for constant $Q$. 

So let us take an ansatz for $C_3$ of the form $C_3=Q(r)\omega\wedge \Lambda$, and find the effective action for $Q$. The minimum will correspond to a stable configuration of $G_4$ flux, supported by the Taub-NUT geometry. Since the fields only depend nontrivially on four dimensions, the M-theory Chern-Simons term will not contribute. The only contribution then comes from the 4-form kinetic term. One can show that
\begin{align}G_4\wedge * G_4=-N^2\left[\ddot{g}^2-U^{-1}(g')^2\right] \frac{dr}{r^2} \wedge d\Omega\wedge  d\theta \wedge dV_6,\label{Mthlag}\end{align} 
in terms of $g=F^2Q$.  The differential equation obtained from \eq{Mthlag} for the eigenmodes $g=g_\omega e^{i\omega t}$ is
\begin{align}\frac{d}{dr}\left(\frac{1}{U r^2}g_\omega'\right)+\frac{\omega^2}{r^2} g_\omega=0.\label{eomth}\end{align}
A solution describing BGHHS charge would be stationary and have a constant $Q$ far away from the hole, so we should set $\omega=0$ above. The most general solution then depends on two constants,
\begin{align}g_0=c_1\left(\frac{Nr^2}{2}+\frac{r^3}{3\lambda^2}\right)+c_2.\end{align}
At large $r$, we have approximately constant $F(r)$. Hence, it is clear that solutions with $c_1\neq0$ do not describe a localized source with finite BGHHS charge at infinity (and actually their energy is divergent). On the other hand, since $U\sim 1/r$ for small $r$, $g$ must go to zero as fast as $r^2$ as  $r\rightarrow0$ for  $G_4$ to actually be smooth near the TN centre. This sets $c_2=0$. 

 Therefore, the boundary conditions forbid stationary solutions, and as a result the pure D6-brane system does not really carry a nontrivial $Q$ charge. Nonzero frequency eigenmodes satisfying \eq{eomth} are dispersive; the TN center reflects incoming waves of $Q$, but is not able to capture any charge.

We might wonder if this behavior is an artifact caused by the fact that the four dimensional dilaton black hole described by the stack of D6-branes is horizonless. The metric and dilaton profile are \cite{Ortin:2004ms}
\begin{align}ds^2=-H^{-1/2} dt^2+H^{1/2}(dr^2+r^2d\Omega),\quad A_0=\pm N\left(H^{-1}-1\right),\quad H=1+2m/r,\label{D6metric}\end{align}
Due to the presence of unstabilized moduli, the solution does bit present an horizon, having a naked singularity instead. This has a simple solution: One can turn the $D6$ system into a genuine black hole by choosing background gauge fields on the brane which induce lower D-brane charges \cite{Ortin:2004ms}. In order to get a Reissner-Nordstrom black hole with a finite horizon, we need to induce D0-, D2-, and D4-brane charge. Turning a fieldstrength for the gauge field along the three independent 2-cycles of $T^6$ will suffice. 

This has a very natural interpretation in M-theory: Since the $D6$ gauge field arises from dimensional reduction of the M-theory 3-form, the above gauge fluxes lift to M-theory backgrounds for $G_4$ of the form $G_4=\omega\wedge \alpha_i$, where $\alpha_i$, $i=1,2,3$ is a basis of independent 2-cycles of the $T^6$. D4 charge is induced simply because there are nontrivial periods of $G_4$; D2-charge is induced by the M-theory Chern-Simons coupling. Finally, D0-charge is induced because the induced M2-charge sits on a magnetic field, and as a result moves along the compact dimension.

We now have an uplift of a charged extremal dyonic BGHHS black hole to a horizonless geometric background in M-theory, yet the equations of motion of the three-form remain the same, because no new contribution comes from the M-theory Chern-Simons term, at least if we ignore gravitational backreaction. Thus, the dispersive behavior of $C_3$ in the black hole background is a genuine feature of the M-theory description of the BGHHS black hole, and not an artifact due to the absence of a horizon in \eq{D6metric}.

It would seem that the only way for the equation of motion coming from \eq{Mthlag} to have a static solution with nonzero constant asymptotics for $Q$ is to allow $G_4$ to diverge at the origin. Then the solution with 
$c_2\neq0$, which corresponds to an asymptotically constant charge $Q=c_2/F^2$. Here $G_4$ vanishes everywhere but at the TN center. This is only consistent if $C_3$ is a gauge transformation of the vacuum, so that
\begin{align}C_3 \in H_3(\mathbb{R}^4-\{0\},\mathbb{Z})\end{align}
describes a pure gauge configuration in singular gauge. Compactness of the gauge group forces $Q$ to be an integer.

To sum up, the M-theory uplift allows us to see explicitly that there is a potential for $C_3$, and that the only smooth stationary configurations have trivial $C_3$.  This is consistent with the conjecture advocated in Section \ref{sec:conj}: The IIA Chern-Simons term 
\begin{align}\int F_8 F_0 B_2\end{align}
becomes upon six-torus compactification a $F_2 B_2 F_0$ coupling. Direct comparison to the M-theory picture is obscured by the somehow cumbersome lift of $F_0$ to M-theory \cite{Hull:1998vy}.

\subsection{Other examples}\label{sec:otherexamples}

Here we collect a few more examples without a detailed discussion:

\medskip

$\bullet$ {\bf Non-geometric fluxes}

The discussion of Chern-Simons terms from internal field strength fluxes in Section \ref{sec:fluxed} clearly generalize to more general fluxes in the compactification, including geometric or non-geometric fluxes (see \cite{Ibanez:2012zz} for review). This is natural, since they are all on equal footing (due to underlying string dualities) from the perspective of the lower-dimensional effective theory.

We now describe an instance of Chern-Simons terms involving nongeometric fluxes, by using T-duality with NSNS 3-form flux. Consider, for instance, type IIB on $T^6$, or on a manifold $X_6$ given by a $T^3$ fibration over a three-dimensional basis $B_3$. In four dimensions, there is a 2-form $C_2$ obtained from dimensional reduction of the ten-dimensional $C_8$. There is no apparent Chern-Simons term involving it in the effective field theory, but this is only because we are ignoring contributions from non-geometric sources. Let us T-dualize three times along $T^3$ in order to make the missing CS term manifest, as follows. In the T-dual type IIA there is a 10d Chern-Simons term $C_5H_3F_2$ which upon dimensional reduction leads to $h \,C_2F_2$ where $h=\int_{T^3} H_3$ and $C_2=\int_{B_3}C_5$. Upon T-duality, $C_5$ goes to $C_8$, while $F_2$ turns into $F_5$, and the NSNS flux becomes a locally non-geometric flux of $q$-type \cite{Shelton:2005cf}. Therefore, in type IIB we indeed have a 10d coupling $qC_8F_5$ which leads to a 4d coupling  $h\,C_2F_2$ upon dimensional reduction on $X_6$. This is the coupling for the 2-form $C_2$ necessary to break its 2-form generalized global symmetry.

 The general lesson is that in string compactifications, whenever there is a BF coupling missing and the global symmetry seems unbroken, it is simply because we sit in a flux-less phase in which the tunable parameter $G_0$ has been naively set to zero. A deeper look reveals that string theory  provides the requested Chern-Simons term to break the global symmetry. In this case, the solution came from the legitimate possibility of nucleating bubbles with a non-vanishing non-geometric flux inside.

\medskip

$\bullet$ {\bf AdS$_5\times S^5$ revisited}

Keeping on with the philosophy of seeking for Chern-Simons terms involving any kind of flux, we can provide a nice description of Chern-Simons couplings breaking the 3-form generalized global symmetry in the $AdS_5\times S^5$ example of Section \ref{sec:holex}. In order to derive it, we again use T-duality, now along the $S^1$ fiber in the $S^5$ when regarded as a circle fibration over $CP_2$ \cite{Duff:1998us}. In the T-dual type IIA on  $AdS_5\times CP^2\times S^1$, there is a 10d Chern-Simons coupling $C_7 H_3 F_0$, which gives a five-dimensional $C_3 H_2 F_0$ term, where $C_3=\int_{CP_2} C_7$, $H_3=\int_{S^1} H_3$. As shown in Section \ref{sec:holex}, we do not actually need this term to break the symmetry - this can be achieved by stringy effects.  Nevertheless, it is also present, allowing us to understand the symmetry breaking in field theory terms. 
Incidentally, it is interesting that it involves the KK $U(1)$ associated to the circle fiber. Since this is part of the $SO(6)$ isometry group of the $S^5$, it would be interesting, but beyond our present scope, to explore the general story associated to the non-abelian structure of this group.

\medskip

$\bullet$ {\bf Heterotic CY compactifications}

Heterotic string theories have non-abelian gauge bosons in 10 dimensions. In either ten, or four dimensions after compactification, one can consider holonomies of the Cartan generators around black objects (7-brane in ten dimensions, string in four), or non-trivial cycles in spacetime. The shift symmetries of these holonomies constitute problematic 1-form generalized global symmetries, which string theory should take care of. The required Chern-Simons terms indeed come from the Green-Schwarz coupling (see \cite{Ibanez:2012zz} for  review)
\begin{align}B_2 \wedge \text{tr}(F^4).\label{gsc}\end{align} 
For instance, in compactifications of the $SO(32)$ heterotic on a CY $X_6$ with standard embedding, the $SO(32)$ gauge group is broken to $SO(26)\times U(1)$. Denoting by $F$ the $U(1)$ fieldstrength, the  coupling \eq{gsc} gives rise to 
\begin{align}\Big(\,\int_{X_6}\text{tr}(F_{SU(3)}^3) \, \Big)\, B_2 \wedge F_{U(1)}\, .\label{gsc2}\end{align} 
The prefactor is non-zero because it equals the holomorphic Euler characteristic of the CY, so we get a four-dimensional coupling $\chi\, B_2\wedge F$, which breaks the symmetr. This can be thought of as a triple CS term by regarding $\chi$ as a slightly exotic version of $F_0$ in this picture. It is quantized, and we can consider four-dimensional domain walls separating regions with different $\chi$: they can correspond to gauge backgrounds changing the gauge bundle topology (thus moving onto a non-standard embedding), or more dramatically to geometric defects interpolating between different CY spaces as we cross the domain wall (via e.g.  conifold transitions). 

Similarly, the dual Green-Schwarz coupling $B_6\wedge\text{Tr}(F\wedge F)$ results in a four-dimensional $\phi F\wedge F$ coupling between the axion dual to $B$ and $F$. Thus we seem to get an axion $\phi$ in four dimensions with both $BF$ and Kaloper-Sorbo like couplings. The low-energy lagrangian
\begin{align}\frac12\vert d\phi-A\vert^2+\phi F\wedge F\end{align}
is not gauge invariant at first sight. However, it is once we take into account the anomalous variation of the four-dimensional chiral fermions; this is just the four-dimensional version of  the usual Green-Schwarz anomaly cancellation.

\medskip

$\bullet$ {\bf 6D $(2,0)$ theories}

Six-dimensional $(2,0)$ theories contain tensor multiplets whose 2-form gauge fields show an associated $2$-form global symmetry. There are no Chern-Simons terms that we can think of for this field, but there is no need for them either, because the symmetry always seems to be gauged. When realizing this theory as the worldvolume theory of a (stack of) M5-branes, the symmetry is gauged by the coupling to the M-theory three-form $C_3$ (one way to see this is to reduce to IIA, where the M5's become D4's and the coupling between the tensor and $C_3$ becomes the usual coupling between $B$ and $F$ in the DBI action). The same happens for the realization in terms of IIA NS5-branes. When the $(2,0)$ theory arises from type IIB in an ALE singularity, the $(2,0)$ B-field arises from reduction of $C_4$ on a basis of normalizable 2-forms of the ALE space. The 3-form also arises from $C_4$, from the decomposition $C_4= B \wedge \omega +C_3 \Lambda$ which is reminiscent of the massive Wilson line examples discussed in \cite{Marchesano:2014mla}.

\section{\texorpdfstring{$3$}{3}-form global symmetries} \label{sec:3forms}
So far we have only considered generalized global symmetries arising from periods of $p$-form gauge fields from $p=0$ to $p=d-2$. These cases can lead to the problematic two-dimensional axion system discussed in Section \ref{sec:unbroken}. The only remaining interesting possibility in four dimensions is that of a $3$-form global symmetry\footnote{A $4$-form global symmetry is also possible in the Euclidean theory. We would have a flat $D_4$ potential with no dynamics (and no charged objects, due to tadpole cancellation), but which might have nontrivial periods on compact 4-manifolds.}. 

This case doesn't immediately fit into our framework, because under compactification to two dimensions it leads to a gauge field, rather than an axion. Additional compactification on a circle leads to an axion in $0+1$ dimensions, that is, the quantum mechanics of a particle on a circle. It is unclear what the problem is with global symmetries in this case, if any at all; we do not know how to relate the $0+1$-dimensional system to any reasonable black hole background, as we did with the $B$-field in Section \ref{sec:hair}. Some solutions, such as the Giddings-Strominger wormhole \cite{Giddings:1987cg}, support nontrivial 3-cycles. If there is a 3-form $C_3$ in our theory, shifts in the period of $C_3$ on such 3-cycle constitute a symmetry of the theory. This is similar to the 1- and 2-form examples, but the similarities seem to end there. Because we are necessarily working in the Euclidean theory now, there doesn't seem to be an immediate problem with the continuous degree of freedom associated to $C_3$: It is just another seemingly harmless zero mode of the euclidean solution. 

This is reminiscent of the situation one encounters when trying to apply the WGC to axions \cite{Rudelius:2014wla,Montero:2015ofa, Brown:2015iha,Brown:2015lia}: One would like to argue that because of the WGC there should be some instanton with an action scaling as $M_P/f$, but the usual trouble-with-remnants justification for the WGC does not work in the Euclidean: There is no natural notion of ``decay'' from one instanton to another (see however \cite{Hebecker:2016dsw} for a recent proposal). 

Because of this, the case for both the WGC to instantons and for our conjecture to 3-forms is weaker than in the other scenarios. In any event, it is interesting exercise, which we carry out in this Section, to explore the implications of the conjecture for 3-form fields (and $d-1$-forms in $d$ dimensions, in general), and to explore a few stringy examples. 

As before, we can either break or gauge the 3-form global symmetry. Gauging  would mean introducing a 4-form gauge potential $D_4$, and modifying the $3$-form kinetic term to
\begin{align}-\frac12\vert F_4-kD_4\vert^2,\end{align}
which would have a tadpole essentially requiring $F_4=0$, thus killing the 3-form itself. This doesn't seem a very interesting possibility. 

Symmetry breaking requires to have a potential term in the effective action for the axion in 0+1 dimensions coming from $C_3$. Such a potential can come from higher-dimensional Chern-Simons terms; for instance, a four-dimensional Kaloper-Sorbo coupling 
\begin{align}\int G_0C_3\wedge d\phi\end{align}
would reduce to a triple interaction
\begin{align} \int G_0 c \dot{\phi} dt,\quad c\sim\int C_3\label{kstriple}\end{align}
in $0+1$ dimensions. Here $\dot{\phi}$ stands for the time derivative of $\phi$. Thus, if $G_0$ does not vanish identically (as an operator), this term breaks the symmetry. 

If the 3-form version of our conjecture applies as well, it implies that all 3-form gauge fields should enjoy the appropriate CS term (which in four dimensions corresponds to the Kaloper-Sorbo coupling above) to break the global symmetry. This Kaloper-Sorbo coupling underlies the construction of axion monodromy models \cite{Marchesano:2014mla}. There is however a subtlety not present in the previous $p$-form symmetries with $p\leq d-2$. Here the parameter $G_0$ is dual to the field strength of another 3-form field $G_4=d\tilde C_3$ in four dimensions. The above coupling can then be understood as a field-dependent kinetic mixing between $F_4$ and $G_4$,
\beq
\int \phi F_4\wedge *G_4\ .
\eeq
The natural question is whether such a coupling also breaks the 3-form global symmetry of $\tilde C_3$, in addition to the one of $C_3$. From this point of view both 3-forms are on equal footing. In fact, one can rewrite this term as as Kaloper-Sorbo coupling for $G_4$, $F_0G_4\phi$, leading to a non-vanishing divergence of the current $G_4$ and breaking the symmetry. Therefore, it is enough if the dual of $C_3$ appears playing the role of $G_0$ in some Kaloper-Sorbo coupling. Support for this conjecture can be found on the four-dimensional effective theories obtained upon dimensional reduction of Type IIA/B in a Calabi-Yau manifold. There, all 3-form fields appear either coupling an axion \`a la Kaloper-Sorbo or playing the role of $G_0$ in another Kaloper-Sorbo coupling \cite{Bielleman:2015ina,Carta:2016ynn}.

This extension of our conjecture may also have implications for the Bousso-Polchinski mechanism \cite{Bousso:2000xa}, which relies on the presence of (many) non-dynamical 3-forms to provide a landscape finely scanning the  cosmological constant value. The conjecture we have presented would imply that there should be a new kind of domain wall in the theory (which changes $G_0$ in \eq{kstriple}), which turns on Kaloper-Sorbo couplings for the Bousso-Polchinski 3-forms. It would be interesting to explore the precise effect these new couplings may have on the standard Bousso-Polchinski mechanism.

These considerations align nicely with the familiar difficulties with realizing the strict Bousso-Polchinski mechanism in string theory. To implement Bousso-Polchinski, we need a decoupled sector with a considerable number of 3-forms. Otherwise, the minimization conditions that stabilize the axions also fix the value of the field strengths 4-forms in the vacuum, which does not leave much freedom to implement the mechanism. However, in practice, the different 3-forms always mix among themselves, and with other fields, as would necessarily be the case if our conjecture was true.

In Section \ref{sec:conj}, we saw that gravity in $d\geq 4$ dimensions would be incompatible with our conjecture. If the $(d-1)$-form version of our conjecture holds, it also excludes $d=3$, since the KK photon would be a 2-dimensional gauge field without Chern-Simons term.  It is worth mentioning that there is a long-standing quest \cite{Witten:2007kt,Gaiotto:2007xh,Gaberdiel:2007ve,Gaiotto:2008jt,Avramis:2007gx, Gaberdiel:2008xb,Maloney:2009ck,Keller:2014xba,Williams:2014rya,Benjamin:2015ria,Benjamin:2016aww} to find a holographic dual to a weakly coupled $d=3$ AdS theory of pure gravity.  The extended version of our conjecture would put it in the Swampland, which seems to align nicely with recent results \cite{Bae:2016yna}. 
Our conjecture doesn't rule out gravity with minimal matter content. One possibility is adding a single axion, which can undergo Scherk-Schwarz compactification as in Section \ref{sec:conj}.

\section{Conclusions}\label{sec:conclus}
The motivational observation of this work is that the usual mechanism of explicit breaking of generalized global symmetries coming from periods of $p$-form potentials via charged particles does not seem to work, at least straightforwardly, for low enough dimensions, e.g. those related to two-dimensional axion systems, where a remnant shift symmetry of the axion seems to survive.

On the face of it, one can take two different viewpoints: Perhaps the presence of global symmetries is not problematic on such low-dimensional systems, or, alternatively, the symmetries must be broken by another mechanism. We have given a particular example, the BGHHS black hole, in which an unbroken symmetry would lead to a remnant problem. Thus, at least in this particular example, there is a very good case for the symmetry to be broken. We have found that this is indeed the case in a consistent embedding of the BGHHS black hole in the $AdS_5/CFT_4$ correspondence. The breaking is due to the contributions of euclidean saddles with the same asymptotics but different topology; from the two-dimensional point of view, these saddles have stringy-sized ``bubbles'' where the effective field theory description breaks down, but which nevertheless break the symmetry.

Even though the bubble that breaks the symmetry in this case is of stringy size, we have also explored bubbles which admit a field theory description as well. These are typically related to triple Chern-Simons terms in the effective field theory, which turn on and off symmetry breaking couplings. We have found these Chern-Simons terms and the associated bubbles to be a generic feature of string compactifications. They are present in every  string compactification we have checked so far, and have varied 10-dimensional origin, ranging from supergravity Chern-Simons terms to non-geometric fluxes and D-brane Chern-Simons terms. Another way to put it is that string theory always seems to allow for both breaking or gauging in different phases of the theory, and there are domain walls which interpolate from one to another. 

We have upgraded our observations to a conjecture, claiming that consistent theories of quantum gravity suffer from a pandemic of Chern-Simons terms. If true, the conjecture further constraints field theories which can be consistently coupled to gravity\footnote{We should remark that the constraints apply to the full theory, and not the low-energy EFT. The symmetry-breaking Chern-Simons terms must be there, but can show up at any scale, much like we expect $B-L$ to be broken or gauged in the UV completion of the Standard Model, but we don't know at which scale the symmetry breaking effects or unhiggsing takes place.}, and puts Einstein-Maxwell+WGC particles, $\mathcal{N}=8$ SUGRA, as well as pure gravity in $d\geq 4$ dimensions in the Swampland. A slight extension of our conjecture to $(d-1)$-forms puts $d=3,4$ pure gravity in the Swampland as well, and creates some tension between a strict Bousso-Polchinski mechanism and quantum gravity. 

One open question is that  we already have  a ``gravitational'' symmetry breaking mechanism in the holographic example, related to the sum over topologies in the holographic description of AdS quantum gravity, so one might wonder if such UV mechanism are more general, and if so why we need the Chern-Simons terms as well. It is unclear to us whether this mechanism works in general, irrespectively of which two-dimensional compactification we consider; in some simple examples, such as a single $U(1)$ on $\mathbb{R}^2\times T^2$, it naively doesn't.  By contrast, the Chern-Simons terms (or more precisely, the associated bubbles) explicitly break the symmetry in any background. We therefore take the presence of the Chern-Simons terms as a general criterion of consistency of the theory, which must be ``ready'' to solve the problems of generalized global symmetries for arbitrary compactifications. Notice that even in this holographic example, we can also find the appropiate Chern-Simons term to break the symmetry.

It would be interesting to gain further insight into this criterion in other holographic setups. A possibly interesting point is that, in the context of the AdS/CFT correspondence, three-dimensional bulk Chern-Simons term can sometimes be related to anomalies of a $CFT_2$ current, so that if we start with a higher current in a higher-dimensional $CFT$ and compactify to $AdS_3$, we necessarily generate bulk Chern-Simons terms, as our conjecture demands. 

In the end, this work has provided a substantial amount of solid support  for our conjecture, coming from diverse examples. Like other conjectures, it might just be that we have only looked at special classes of string vacua. However, the conclusion so far seems strong enough to support the proposal of this criterion as a generic feature of the stringy landscape/swampland. We expect further work in this direction to lead to progress in the understanding of generalized global symmetries, and of the role of Chern-Simons terms in string theory. 

\subsubsection*{Acknowledgements}
We thank William Cottrell, Thomas Grimm, Arthur Hebecker, Fernando Quevedo, Diego Regalado, Pablo Soler, and Gianluca Zoccarato for helpful discussions and comments. MM is supported by a Postdoctoral fellowship from ITF, Utrecht. AU is supported by the grants FPA2015-65480-P and SEV-2012-0249 of the ``Centro de Excelencia Severo Ochoa" Programme from the Spanish Ministry, and the ERC Advanced Grant SPLE under contract ERC-2012-ADG-20120216-320421. IV is supported by  a grant from the Max Planck Society. MM thanks the Institute for Advanced Studies at the Hong Kong University of Science and Technology for hospitality while this work was completed. MM and IV thank the DESY workshop ``New Ideas in String Phenomenology'' for a stimulating atmosphere which led to several fruitful discussions.

\appendix

\section{Confinement of strings}\label{app:polyakov}
It is possible to emulate the original computation in \cite{Polyakov:1976fu} showing confinement in $(2+1)$ to display the same phenomenon for strings in $(3+1)$. For simplicity, we ignore the effects of gravity for the time being. In the euclidean theory, confinement will be shown by the behavior of the Wilson loop
\begin{align}W[S]=\left\langle\exp\left(i\int_{S} B\right)\right\rangle.\end{align}
If we take the surface $S$ to be $C\times[0,T]$,  where $C$ is some closed curve in three dimensions, then for very large $T$ this represents the amplitude for a nondynamical string to nucleate along $C$ only to annihilate after a very long time $T$. In this limit, the amplitude is a tunneling effect and is approximately $e^{-E[C] T}$, where $E[C]$ is the energy of the configuration. However, one may also write
\begin{align}W[S]=\left\langle\exp\left(i\int_{V} *d\phi\right)\right\rangle.\end{align}
where $S=\partial V$. 

The term is just the partition function for an axion. If the dilute-gas approximation for the instantons is valid (the conclusion can be modified significantly when it is not), we have an axion with modified action
\begin{align}S=\int \frac12 d\phi \wedge d\phi+\Lambda(1-\cos(\phi/f)+d\phi\wedge*\text{P.D}(V)\label{modact}\end{align}
where $\text{P.D}(V)$ is the Poincar\'e dual of $V$. It is straightforward to evaluate this around the solution to the equations of motion. If we take the curve $C$ to lie in the $xy$ plane, then 
\begin{align}\text{P.D}(V)=\chi(S)\delta(z) dz\end{align}
where $S$ is the region bounded by $C$ and $\chi(S)$ is its characteristic function. Then, the equations of motion are
\begin{align}\nabla^2\phi=-\Lambda\sin\left(\frac{\phi}{f}\right)+\delta'(z) \chi(S).\end{align}
Far from the boundaries, this is a one-dimensional equation with solution
\begin{align}\phi(z)=4\, \text{sign}(z)\arctan\left(e^{-m\vert z\vert}\right)\label{ssssdd}\end{align}
where $m=\sqrt{\Lambda/f^2}$. Plugging back into the path integral, fluctuations around this solution become gaussian and exactly computable, giving an overall factor. The term coming from the classical action will give a contribution of the form $\gamma \text{Vol}(V)$, where $\gamma$ is computed by evaluating the 1-dimensional version of the action \eq{modact} plugging \eq{ssssdd}. 

This area (or rather, volume) behavior of the Wilson loop shows confinement of the strings. As we try to stretch a string to a length enough to lasso the black hole, an anti-string will be nucleated.The anti-string will provide an equal and opposite contribution to the Aharonov-Bohm phase, rendering the charge $Q$ unobservable in practice.

\section{Self-dual fluxes in Taub-NUT space}\label{app:TNdetails}
In this brief appendix we will derive the expression for the self-dual form of Taub-NUT space. For convenience, we reproduce again the Taub-NUT metric
\begin{align} ds^2=U(d\vec{x}\cdot d\vec{x})+U^{-1}(d\theta + M)^2,\quad U=\frac{1}{\lambda^2}+\frac{N}{r}.\label{tnapp}\end{align}

First we will find a potential for the Taub-NUT self dual 2-form $\omega=d\Lambda$, with an ansatz (inspired by the $N=1$ case) of the form
\begin{align} \Lambda=F(r)(d\theta+M).\label{sdgshda}\end{align}
A tetrad basis for the metric \eq{tnapp} is
\begin{align} e_1=\sqrt{U} dr,\ e_2=\sqrt{U} rd\theta, \ e_3=\sqrt{U} r\sin\theta d\phi, \ e_4=\frac{1}{\sqrt{U}}(d\theta+M).\end{align}
The exterior derivative of \eq{sdgshda} is
\begin{align}\omega=d\Lambda=F' e_1\wedge e_4- \frac{NF}{Ur^2}e_2\wedge e_3,\end{align}
since 
\begin{align}dM=*_{3D}dU=-\frac{N}{r^2}*_{3D} dr=-Nd\Omega.\end{align}
Anti-self-duality then amounts to 
\begin{align}F'=\frac{NF}{Ur^2}\quad\Rightarrow\quad F=\frac{r}{r+N\lambda^2}.\end{align}

\section{ 2d instantons and Chern-Simons terms}\label{app:fat}

We now discuss a heterotic example which perhaps sheds some light on the relationship between Chern-Simons terms and charged objects: Consider the same heterotic CY compactification as Section \ref{sec:examples}, but focus on a different  4d $B$-field, coming from wrapping $B_6$ on a 4-cycle $A_4$.  The triple Chern-Simons comes from dimensional reduction of $B_6\text{tr}F^2$, where $G_0$ is the component of the first Chern class of the gauge bundle along the 2-cycle $A_2$ dual to $A_4$. 

The 10d Chern-Simons term $B_6\text{tr}F^2$ also endows heterotic gauge instantons with NS5-brane charge \cite{Witten:1995gx,Ovrut:2000qi}. The NS5-brane can be regarded as a point-like gauge instanton of vanishing size or, put in another way, the $B_6\text{tr}F^2$ term allows NS5-branes to fatten. The strings coupled to the 4d $B$ field are just NS5-branes wrapped on $A_4$; if we further compactify to two dimensions, they become instantons for $b=\int B$. As usual, their action is IR divergent. However, we may fatten the NS5 brane to a $SO(32)$ instanton of radius $\rho$. If we are not concerned with solving the Euclidean equations of motion, but only with finding a finite action deformation of the instanton, we are allowed to replace the NS5 with a $U(1)$ gauge field configuration with
\begin{align}F=\xi(r) (dV_2+ \omega_{A_2}),\label{NS5fat}\end{align}
where $\omega_{A_2}$ is the Poincar\'e dual to $A_{2}$, and $dV_2$ is the volume element of the two-dimensional noncompact factor. The smooth function function $\xi(r)$ vanishes for $r>\rho$ and is such that $\int F\wedge F$ has the instanton number of a NS5.

So far, we have only fattened the two-dimensional instanton - but it still has a IR-divergent action.  Consider adding now a anti NS5, on top of the NS5, which fattens to
\begin{align}F=\xi(r) (-dV_2+ \omega_{A_2}).\label{antiNS5fat}\end{align}
The configuration now consists of an instanton- anti instanton pair, and its action is no longer divergent. However, the gauge fields \eq{NS5fat} and \eq{antiNS5fat} do not cancel out: they leave a component along $\omega_{A_2}$ or, in other words,
\begin{align}G_0=2\xi(r),\end{align}
which is precisely one of the bubbles introduced in Section \ref{sec:break}. At $r\sim\rho$, we have $dG_0=d*G_2\neq0$ so that the boundary of the bubble is a particle charged electrically under $G_2$, as discussed there. In other words, the symmetry-breaking bubble can be constructed by fattening and deforming a finite action instanton- anti instanton configuration. In Section \ref{sec:unbroken}, we argued that charged objects alone are not enough to break the symmetry. Here, we see that in the presence of Chern-Simons terms they can fatten, which in turn allows them to break the two-dimensional symmetry.

This behavior may seem particular to the heterotic case, but we expect it to hold at least also for type II examples. The D-brane instantons one gets may fatten due to brane polarization \cite{Myers:1999ps}; it is possible to overlap a polarized D-brane instanton-anti instanton pair to construct the bubble in the same way. 

\bibliographystyle{jhep}
\bibliography{refs-GGS-2}
 \end{document}